\documentclass[a4paper,11pt]{article}

\usepackage[english]{babel}
\usepackage{setspace}
\usepackage[a4paper,tmargin=3.5truecm,bmargin=4truecm,rmargin=2.5truecm,
lmargin=3.2truecm,twoside,verbose=true]{geometry}
\usepackage{cancel,graphicx}
\usepackage{hyperref}
\usepackage{amsmath,amssymb,slashed}

\usepackage{amsmath,amssymb}
\usepackage[amsmath, hyperref, thmmarks]{ntheorem}
\numberwithin{equation}{section}

\allowdisplaybreaks[1]

\newcommand{\be}{\begin{equation}}
\newcommand{\ee}{\end{equation}}
\def\beqa{\begin{eqnarray}}
\def\eeqa{\end{eqnarray}}
\def\nn{\nonumber}
\newcommand{\eqn}[1]{(\ref{#1})}

\newcommand\bbbone{\mathbb{I}}
\newcommand\Rl{\mathbb{R}^3_\lambda}

\newcommand\caZ{{\mathcal Z}}

\newcommand\gR{{\mathbb R}}
\def\gC{{\mathbb C}}

\def\gR{{\mathbb R}}
\def\N{{\mathbb N}}

\newcommand\algebA{{\mathbb A}}

\newcommand\modul{{\mathbb M}}

\newcommand\calg{{{\cal{G}}}}

\newcommand\DER{{\text{\textup{Der}}}}

\newcommand\del{{\partial}}

\DeclareMathOperator{\tr}{Tr}

\newcommand\dd{{\text{\textup{d}}}}

\newcommand{\cg}[6]{
  \left(
  \begin{array}{cc|c}
  #1 & #3 & #5 \\
  #2 & #4 & #6
  \end{array}
  \right)
}
\newcommand{\wign}[6]{
\left(
  \begin{array}{ccc}
  #1 & #3 & #5 \\
  #2 & #4 & #6
  \end{array}
  \right)
}

\renewenvironment{thebibliography}[1]
         {\section*{References}\frenchspacing\small
          \begin{list}{[\arabic{enumi}]}
         {\usecounter{enumi}\parsep=2pt\topsep 0pt
         \settowidth{\labelwidth}{[#1]}
         \leftmargin=\labelwidth\advance\leftmargin\labelsep
         \rightmargin=0pt\itemsep=1pt\sloppy}}{\end{list}}

\theoremsymbol{}
\theorembodyfont{\slshape}
\theoremheaderfont{\normalfont\bfseries}
\theoremseparator{}

\theorembodyfont{\upshape}
\theoremsymbol{\ensuremath{\blacklozenge}}

\theoremstyle{nonumberplain}
\theoremheaderfont{\scshape}
\theorembodyfont{\normalfont}
\theoremsymbol{\ensuremath{\blacksquare}}

\qedsymbol{\ensuremath{_\blacksquare}}
\theoremclass{LaTeX}

\title{Quantum gauge theories on noncommutative 3-d space}
\author{Antoine G\'er\'e$^a$, Patrizia Vitale $^{b,c}$ and Jean-Christophe Wallet$^d$}

\begin{document}

\date{}
\maketitle
\vspace*{-1cm}

\begin{center}
\textit{$^a$Dipartimento di Matematica, Universit\`a di Genova\\Via Dodecaneso, 35, I-16146 Genova, Italy} \\
\textit{$^b$Dipartimento di Fisica
Universit\`a di Napoli Federico II}  \\
\textit{$^c$INFN, Sezione di Napoli, Via Cintia 80126 Napoli, Italy}\\
\smallskip
\textit{$^d$Laboratoire de Physique Th\'eorique, B\^at.\ 210\\
CNRS and Universit\'e Paris-Sud 11,  91405 Orsay Cedex, France}\\
\bigskip
 e-mail:
\texttt{gere@dima.unige.it,patrizia.vitale@na.infn.it, jean-christophe.wallet@th.u-psud.fr}\\[1ex]

\end{center}

\begin{abstract}
We consider a  class of gauge invariant models on the noncommutative space $\mathbb{R}^3_\lambda$, a deformation of the algebra of functions on $\mathbb{R}^3$. Focusing on massless models with no linear $A_i$ dependence, we obtain noncommutative gauge models for which the computation of the propagator can be done in a convenient gauge. We find that the infrared singularity of the massless propagator disappears in the computation of the correlation functions. We show that massless gauge invariant models on $\mathbb{R}^3_\lambda$ have quantum instabilities of the vacuum, signaled by the occurrence of non vanishing 1-point functions for some but not all of the components of the gauge potential. The tadpole contribution to the effective action cannot be interpreted as a standard $\sigma$-term. Its global symmetry does not fit with the one of the classical action, reminiscent of an explicit global symmetry breaking term.
\bigskip

\end{abstract}

\pagebreak
\section{Introduction}\label{intro}

%The idea of noncommutative space-time  goes back to the beginning of quantum mechanics. Already Heisenberg and then Snyder conjectured that it %could help solving the problem of UV divergences in quantum field theory. 
%A modern point of view on the necessity of abandoning the notion of a classical space-time and commuting coordinates, which are likely to become  meaningless at  Planck scale, can be found in  \cite{Doplich1}.  
Noncommutative field theories 
(NCFT) 
%originating from noncommutative geometry \cite{Connes1} 
were formulated in  their modern form in the mid 80's first within string field theory \cite{witt1}, followed by pioneering models on the fuzzy spheres and almost commutative geometry \cite{DKM}, \cite{fuzzy1}, while some types of NCFT on Moyal space $\mathbb{R}^4_\theta$ were identified as possible effective regimes of string theory at the end of the 90's \cite{Schomerus}, attracting a lot of interest. For an accurate description of Moyal spaces, see \cite{pepejoe}.  Reviews on Moyal NCFT may be found in e.g.~\cite{Szabo:2001}.\par

Recently, scalar field theories on the noncommutative space $\mathbb{R}^3_\lambda$, a deformation of $\mathbb{R}^3$ which preserves rotation invariance, have been studied in \cite{vit-wal-12}. These appear  to have a mild perturbative behavior and are (very likely) free of ultraviolet/infrared (UV/IR) mixing. In this respect, one may expect a more favorable situation for the gauge theories on $\mathbb{R}^3_\lambda$ than for those on $\mathbb{R}^4_\theta$ whose present status is recalled below. The purpose of this paper is to examine the interesting case of gauge invariant theories that can be built on $\mathbb{R}^3_\lambda$, focusing on some quantum properties. The space $\mathbb{R}^3_\lambda$, which may by viewed as a subalgebra of $\mathbb{R}^4_\theta$,  has been first introduced in \cite{Hammaa} and generalized in \cite{selene}. The use of the canonical matrix basis introduced in \cite{vit-wal-12} (see also \cite{duflo}) renders the computation tractable, avoiding the complexity of a direct calculation in coordinates space, the star product of $\mathbb{R}^3_\lambda$ being of Lie algebra type.\par

Most of the NCFT are non local. Unless one deals with a finite noncommutative geometry, their perturbative renormalization is difficult, as it is the case   for Moyal spaces. This comes mainly from the UV/IR mixing which shows up already in the real-valued $\varphi^4$ model and also in gauge models on $\mathbb{R}^4_\theta$  of Yang-Mills  type, such that the one considered in \cite{Minwalla:1999px}.   A first solution to this problem is provided by the popular Grosse-Wulkenhaar model on $\mathbb{R}^{2n}_\theta$, $n=1,2$,   which is renormalizable to all orders \cite{Grosse:2003aj-pc} and is moreover very likely to be non-perturbatively solvable \cite{harald-raimar}. Various aspects of the Grosse-Wulkenhaar model have been examined, among which classical and/or geometrical ones, as well as $2$-d fermionic extensions \cite{bwh-1}-\cite{vtw}. The initial success of the Grosse-Wulkenhaar model triggered attempts to extend its features to a gauge theoretical framework. In this spirit, a gauge invariant model obtained either by effective action computation or by heat kernel methods was proposed in~\cite{Wallet:2007c}. This model appears to be linked to a particular type of spectral triple \cite{Grosse:2007jy} whose relationship to the Moyal geometries has been analyzed in \cite{Wallet:2011aa}. Unfortunately, its complicated vacuum structure explored in \cite{GWW2} forbids the use of any standard perturbative treatment. Nevertheless, this technical obstruction can be overcome in the case of $\mathbb{R}^2_\theta$ at least for a particular class of vacuum configurations, once the gauge invariant model is treated as a matrix model \cite{MVW13} showing incidentally  a relationship with an extension of a 6-vertex model. Whether or not this can be actually extended to $\mathbb{R}^4_\theta$ is not known at the present time. Other approaches attempting to avoid the vacuum problem or proposing an alternative approach  have appeared, see \cite{Blaschke:2009c}-\cite{Blaschke:2010ck}. The noncommutative differential calculus related to  these gauge models has been explored in \cite{Wallet:2008bq}-\cite{Wallet:2008b}. The construction of a renormalizable gauge theory on $\mathbb{R}^4_\theta$ is still unsolved.\par

In this paper, we consider a  class of gauge invariant models built on the noncommutative space $\mathbb{R}^3_\lambda$, stemming from a natural differential calculus based on a Lie algebra of derivations of $\mathbb{R}^3_\lambda$ \cite{Marmo:2004re}. In order to introduce the connection, we thus use the standard noncommutative analog of the Koszul notion of connection \cite{Wallet:2008bq, Wallet:2011cmj, Wallet:2008b}. Related squared curvature terms into the functional action yield generally  mass terms for the gauge potential $A_i$. In order to mimic salient classical features of commutative Yang-Mills theory, we focus on models which are massless and with no linear $A_i$ dependence. This yields a class of noncommutative gauge models for which the propagator can be computed in a suitable gauge. This latter may be viewed as an analog of the covariant gauges used within commutative gauge theories. Working in this gauge, we find  that the infrared singularity of the massless propagator disappears from the computation of the correlation functions. We show that massless gauge invariant models on $\mathbb{R}^3_\lambda$ have quantum instabilities of the vacuum, signaled by the occurrence of non vanishing tadpole (1-point) functions for some but not all of the components of the gauge potential. The tadpole contribution to the effective action cannot be interpreted as a standard $\sigma$-term. Its global symmetry does not fit with the one of the classical action which is reminiscent of an explicit global symmetry breaking term.\par 
Interestingly, the action for the gauge models described above, when reduced to a single fuzzy sphere of the ``foliation'' of $\mathbb{R}^3_\lambda$,  yields  the Alekseeev-Recknagel-Schomerus action \cite{ARS00}, a natural gauge action on the fuzzy sphere emerging in the context of string theory as the low energy action for brane dynamics on $S^3$, which is combination of a Yang-Mills and a Chern-Simons-like term. 

In  section \ref{section2}, we collect the relevant properties of the noncommutative differential calculus on $\mathbb{R}^3_\lambda$, underlying the above gauge theories,  as well as the related matrix basis used in this paper. In section \ref{section3}, the construction of the classical gauge invariant action is discussed and the computation of the propagator after BRST gauge-fixing is presented. In section \ref{section4}, the computation of the tadpole functions is given in detail. In section \ref{discussion}, we discuss the results. Some useful related technical material on the perturbative expansion and on the computation of a UV limit is collected respectively in appendix \ref{appendix1} and appendix \ref{appendix2}

\section{\texorpdfstring{Differential calculus on $\mathbb{R}^3_\lambda$ and Yang-Mills action}{Diff calcul}}\label{section2}

\subsection{Derivation based differential calculus in a nutshell}\label{subsection21}

To make the discussion self-contained, we recall briefly the relevant features of the noncommutative differential geometric set-up underlying the present work. 

Let $\algebA$ be an associative $*$-algebra
%with unit $\bbbone$ and 
with center $\caZ(\algebA)$. Let $\DER(\algebA)$ be a Lie algebra of derivations of $\algebA$ with Lie bracket defined by $[X,Y]a:=(XY-YX)a$, $\forall X,Y\in\DER(\algebA)$, $\forall a\in\algebA$, which is only  a module over $\caZ(\algebA)$. Real derivations satisfy $(X(a))^\dag=X(a^\dag)$, $\forall a\in\algebA$. In the noncommutative case derivations are not a module over $\algebA$, therefore for them to be independent  we have to require that they be``sufficient", that is, only elements in the center of $\algebA$ are annihilated by all derivations.  On the other hand, 
The differential calculus based on the derivations of an algebra $\algebA$, introduced  long ago \cite{segal, dbv1, marmolandi, dbv2},  is a generalization of the de Rham differential calculus in which the derivations play the role of the vector fields. For mathematical details and applications to NCFT, we refer the reader to  \cite{ Marmo:2004re, Wallet:2008bq, Wallet:2011cmj, Wallet:2008b}. We just recall here the definition of exterior derivative, since we shall explicitly need  it in the article: 
%Recall that any Lie subalgebra $\calg$ of $\DER(\algebA)$, still a module over ${\cal{Z}}(\algebA)$, generates a differential calculus, with $\mathbb{N}$-graded differential algebra $({\Omega}^\bullet_{\calg}(\algebA) =
%\textstyle \bigoplus_{n \geq 0} {\Omega}^n_{\calg}(\algebA),\ \times,\ d)$, where ${\Omega}^n_{\calg}(\algebA)$ is the space of ${\cal{Z}}(\algebA)$-n-linear antisymmetric maps $\omega:\calg^n\to\algebA$ and the inner %product $\times$ over ${\Omega}^\bullet_{\calg}(\algebA)$ and 
$
d:{\Omega}^n_{\DER}(\algebA)\to{\Omega}^{n+1}_{\DER}(\algebA)$ is defined for any 
$\omega\in{\Omega}^p_{\DER}(\algebA)$,
% $\eta\in{\Omega}^q_{\calg}(\algebA)$ 
by
%\begin{eqnarray}
%(\omega\times\eta)(X_1,..., X_{p+q}) &=&
 %\frac{1}{p!q!} \sum_{\sigma\in \kS_{p+q}} (-1)^{{\textup{sign}}(\sigma)}
%\omega(X_{\sigma(1)},..., X_{\sigma(p)}) 
%\eta(X_{\sigma(p+1)},..., 
%X_{\sigma(p+q)})\nonumber\\
%
\begin{eqnarray*}
d\omega(X_1,..., X_{p+1})& :=& \sum_{i=1}^{p+1} (-1)^{i+1} X_i \omega( X_1,..\vee_i.., X_{p+1}) \\
&+&  \sum_{1\leq i < j \leq p+1} (-1)^{i+j} \omega( [X_i, X_j],..\vee_i..\vee_j.., X_{p+1}) \label{eq:koszul},
\end{eqnarray*}
($\vee_i$ means that the argument $i$ is omitted) and ${\Omega}^0_{\DER}(\algebA)=\algebA$. \par 

We then consider a natural noncommutative extension of the notion of connection, as introduced in \cite{dbv1} which uses (finite projective) right-modules over $\algebA$, somewhat similar to the Koszul connections of the commutative case. Let $\modul$ be a right-module over $\algebA$. A connection on $\modul$ can be conveniently defined %\cite{dbv1, dbv2, Wallet:2008bq} 
by a linear map ${\nabla} :  \DER(\algebA)\times  \modul \rightarrow \modul$ satisfying
\begin{equation}
{\nabla}_X (m a) = mX( a) + {\nabla}_X (m) a,\ {\nabla}_{fX}( m) = f{\nabla}_X (m),\ 
{\nabla}_{X + Y} (m) = {\nabla}_X (m) + {\nabla}_Y (m) \label{connect}
\end{equation}
for any $X,Y \in \DER(\algebA)$, $a \in \algebA$, $m \in \modul$, $f \in \caZ(\algebA)$.  Hermitian connections, used in this paper, satisfy  for any {\it{real}} derivation $X \in \DER(\algebA)$
\begin{equation}
 X(h(m_1,m_2))=h(\nabla_X(m_1),m_2)+h(m_1,\nabla_X(m_2)), \forall m_1,m_2\in\modul,\label{hermitconnect}
\end{equation}
where $h:\modul\otimes\modul\to\algebA$ denotes a Hermitian structure{\footnote{Recall that a Hermitian structure is a sesquilinear map, $h:\modul\otimes\modul\to\algebA$, such that $h(m_1,m_2)^\dag=h(m_2,m_1)$, $h(ma_1,ma_2)=a_1^\dag h(m1,m2)a_2$, $\forall m_1,m_2\in\modul,\ \forall a_1,a_2\in\algebA$.}} on $\algebA$. The curvature is the linear map $F(X, Y) : \modul \rightarrow \modul$ defined by
\begin{equation}
 F(X, Y) m = [ {\nabla}_X,{\nabla}_Y ] m - {\nabla}_{[X, Y]}m,\ \forall X, Y \in \DER(\algebA)\label{courgene}.
\end{equation}

The group of gauge transformations   of $\modul$, ${\cal{U}}(\modul)$, is   defined \cite{Wallet:2008bq} as the group of automorphisms of $\modul$ compatible both with the structure of right $\algebA$-module and the Hermitian structure, i.e $g(ma)=g(m)a,\ h(g(m_1),g(m_2))=h(m_1,m_2)$, $\forall g\in{\cal{U}}(\modul)$, $\forall a\in\algebA$, $\forall m_1,m_2\in\modul$. For any $g\in{\cal{U}}(\modul)$, the gauge transformations are
\begin{eqnarray}
{\nabla}^g_X&:&\modul\to\modul,\ {\nabla}^g_X = g^{-1}\circ {\nabla}_X \circ g\label{gaugeconnect}\\
F(X,Y)^g&:&\modul\to\modul,\ F(X,Y)^g=g^{-1}\circ F(X,Y) \circ g\label{gaugecurv}.
\end{eqnarray}

Since we want to generalize a gauge theory with structure  group  $U(1)$ -electrodynamics-  the relevant vector bundle in the commutative case  is a complex line bundle. This is generalized by means of a one-dimensional module $\modul=\gC \otimes \algebA$. As Hermitian structure  we choose $h(a_1,a_2)=a_1^\dag a_2$ and take real derivations. Then a Hermitian connection is entirely determined \cite{Wallet:2008bq} by its action on the one-dimensional basis $\nabla_X({\mathbb{I}})$. We have  $ \nabla_X(a)= \nabla_X({\mathbb{I}}) a + X(a)$,with  $\nabla_X({\mathbb{I}})^\dag=-\nabla_X({\mathbb{I}})$. This defines in turn the 1-form connection $A$  by  means of 
\begin{equation}
A:X\to A(X):=\nabla_X({\mathbb{I}}), \;\; \forall X\in\DER(\algebA)
\end{equation}
 The group of unitary gauge transformations ${\cal{U}}(\algebA)$ is the group of unitary elements of $\algebA$, acting multiplicatively on the left of $\algebA$. Then,  Eqs. \eqref{gaugeconnect}, \eqref{gaugecurv} yield
\begin{equation}
\nabla_X({\mathbb{I}})^g=g^\dag\nabla_X({\mathbb{I}})g+g^\dag X(g),\ F(X,Y)^g=g^\dag F(X,Y)g,\ \forall X, Y\in\calg,\ \forall a\in\algebA\label{gaugetransfinal}
\end{equation}
for any unitary $g\in\algebA$. \par 

We shall be concerned with inner derivations, that is $X\in \DER(\algebA)$ such that their action on $\algebA$ may be written as a $\star$-
commutator or: $X( a)= [f_X, a]_\star$, for some $f_X\in \algebA$. Let us assume that there exists a fundamental one-form $\eta\in \Omega^1_{\DER}(\algebA)$, such that  
\begin{equation}
X(a)\equiv da (X)= [\eta(X),a], \forall a\in \algebA \label{etaX}
\end{equation}
 with  $\eta(X)= f_X\in \algebA$. Then it can be shown (cfr.\cite{Wallet:2008bq}) that the following maps
\begin{equation}
  \nabla^{inv}_X(a)=X(a)-\eta(X) a=-a\eta(X),\ {\cal{A}}(X):=\nabla_X-\nabla^{inv}_X=A(X)+\eta(X)\label{canonics}
\end{equation}
define respectively a gauge-invariant connection $\nabla^{inv}$,  which we shall refer to as canonical connection,  and a gauge covariant 1-form ${\cal{A}}$ (that is  $\cal{A}$  verifies ${\cal{A}}(X) ^g=g^\dag{\cal{A}}(X)g$). 

\noindent For any $X,Y\in\DER(\algebA)$, the curvature of a given connection $A$, defined in Eq. \eqref{courgene},  may be re-expressed in terms of the tensor form $\mathcal{A}$ as 
\begin{equation}
F{(X,Y)}=([{\cal{A}}(X),{\cal{A}}(Y)]-{\cal{A}}{[X,Y]})-([\eta(X),\eta(Y)]-\eta([X,Y]))\label{courburebis}.
\end{equation}
Moreover it can be  verified that the curvature of the canonical connection satisfies
 \begin{equation}
 F^{inv}{(X,Y)}=\eta([X,Y])-[\eta(X),\eta(Y)]\;\; \in\caZ(\algebA). \label{finv}
\end{equation}
\subsection{\texorpdfstring{The algebra $\mathbb{R}^3_\lambda$}{Diff calcul}}\label{subsection22}

Let us now consider the case $\algebA=\Rl$, a deformation of the algebra of functions on $\mathbb{R}^3$ introduced in \cite{Hammaa} and further studied in \cite{selene}, \cite{vit-wal-12}, \cite{duflo}.  
Denoting by $(x^{i=1,2,3})$ the coordinate functions on $\mathbb{R}^3$, the associative noncommutative product of the algebra is so defined 
\begin{equation}
\phi\star \psi \,(x)= \exp\left[\frac{\lambda}{2}\left(\delta^{ij} x^0+ i \epsilon^{ij}_k x^k \right)\frac{\del}{\del u^i}\frac{\del}{\del v^j}\right] \phi(u) \psi(v)|_{u=v=x} \label{starsu2}
\end{equation}
where $\lambda$ is the  noncommutative parameter of length dimension 1 and $x^0$ a fourth coordinate function, the radius,  defined in terms of the commutative product of  the other three, $(x^0)^2= \sum_i (x^i)^2$. It can be verified  that $x^0$ $\star$-commutes with all elements of the algebra. The star product \eqref{starsu2} implies for coordinate functions
 \begin{eqnarray}
x^i\star x^j&=& x^i x^j+ \frac{\lambda}{2} \left(x^0 \delta^{ij} + i \epsilon^{ij}_k x^k\right);\ x^0\star x^i = x^i\star x^0 = x^0 x^i + \frac{\lambda}{2} x^i;\\
x^0\star x^0&=&(x^0)^{*2}=x^0(x^0+\frac{\lambda}{2}) = \sum_{i=1}^3 x^i\star x^i- \lambda x^0 . \label{xx0*2}
 \end{eqnarray}
from which one obtains
\begin{equation}
 [x^i,x^j]_\star=i \lambda \epsilon^{ij}_k  x^k \;\;\; [x^0,x^j]_\star= 0\label{commsu2}.
\end{equation}
More details on the derivation of the star-product \eqref{starsu2} and the definition of the algebra may be found in appendix \ref{appendix0}. 

Here we just recall that the algebra $\gR^3_\lambda$ has been obtained as a sub-algebra of the Wick-Voros algebra $\gR^4_\theta$. Such an identification 
 has a geometric counterpart in the commutative setting, where the Kustaanheimo-Stiefel (KS) map \cite{KS} can be used.  We review in the following this classical derivation because it allows  the definition of an integral and differential calculus which are easily generalized to the noncommutative case. The discussion below is taken from \cite{V14}.

 The main idea is the observation that $\gR^3-\{0\}$ and $\gR^4-\{0\}$ may be given the structure of trivial bundles over spheres, being $\gR^3-\{0\}\simeq S^2\times\gR^+$ and $\gR^4-\{0\}\simeq S^2\times \gR^+$. Then one may use the well known Hopf fibration $\pi_H : S^3\rightarrow S^2$, with the identification of $S^3$ with $SU(2)$,
 \be
 \pi_H: s\in SU(2)\rightarrow \vec x \in S^2, \;\; : s\sigma_3 s^{-1}= x^i \sigma_i
 \ee
 where $s= y_0\sigma_0+ i y_i \sigma_i$ and $y_\mu$ are real coordinates on $\gR^4$ such that $y_\mu y^\mu=1$. Now one may extend (not uniquely)  the Hopf  map to $\gR^4-\{0\}\rightarrow \gR^3-\{0\}$, relaxing the radius  constraint so that $y_\mu y^\mu=R^2$, with $R\in \gR^+$. On introducing $g= R s$ we define
 \be
 \pi_{KS}: g \in\gR^4-\{0\}\rightarrow \vec x\in \gR^3-\{0\}, \;\;\; x^k\sigma_k=g \sigma_3 g^\dag=  R^2 s\sigma_3 s^{-1}. \label{KS}
 \ee
 One can easily verify that this map gives back relations \eqn{xmu} up to a factor of $2$, with $z_1= \frac{1}{\sqrt{2}} (y_0+i y_3)$, $z_2=\frac{1}{\sqrt{2}}(y_1+i y_2)$ and the identification
 \be x_0=\frac{ R^2}{4}. \label{rR}
 \ee
 The KS fibration may be used to define the derivations for the algebra of functions $\mathcal{F}(\gR^3-\{0\})$ as projections of the derivations of $\mathcal{F}(\gR^4-\{0\})$ \cite{DMV05}. We shall see that this procedure can be generalized to  the noncommutative setting. Moreover, with the introduction of the matrix basis, the restriction to $\gR^3-\{0\}$ may be removed, since we shall see that our matrix basis is well defined in $0\in \gR^3$ as well.

Let us shortly review  the matrix basis adapted to $\Rl$ constructed in \cite{vit-wal-12, duflo}. More details on the derivation may be found in appendix \ref{appendix0}, while here we just state the results. 
The basis elements are represented by 
\begin{equation}
v^j_{m\tilde m}(x)= \frac{ e^{-2\frac{ x_0}{\lambda}}}{ \lambda^{2j}}  \frac{(x_0+x_3)^{j+m}
(x_0-x_3)^{j-\tilde m}\; (x_1-i x_2)^{\tilde m -m}
}{\sqrt{(j+m)!(j-m)! (j+\tilde m)!(j-\tilde m)! }} \label{xmatrixbasis}
\end{equation}
with $j\in\frac{N}{2},  -j\le\;m,\tilde m\le j$.
Elements of the algebra are thus represented by
\be\phi(x)=\sum_{j\in \frac{\N}{2}}\sum_{m,\tilde m=-j}^j \phi^j_{m\tilde m} v^j_{m\tilde m}(x)  \;\;\; \phi^j_{m\tilde m}\in \gC\label{funexp}
\ee
Let us notice that the functions expansion \eqn{funexp} is well behaved in $x=0$, it being $\lim_{x\rightarrow 0}\phi(x) = \phi^0_{00}$ .  Thus,  $v^{j}_{m\tilde m}(x)$ provides a basis for the commutative (upon redefining $\tilde x= x/\lambda$), and noncommutative algebras of functions on the whole $\gR^3$, analogously to the Moyal matrix basis \cite{pepejoe},  under usual regularity assumptions for  the sequence of the coefficients $\{\phi^j_{m\tilde m}\}$.
The star product \eqn{starsu2} applied to the basis elements acquires the simple form
\be
v^j_{m\tilde m}\star v^{\tilde\jmath}_{n \tilde n}(x)=\delta^{j \tilde\jmath}\delta_{\tilde m
n}v^j_{m \tilde n}(x) \label{matrixprod}
\ee
Then, the star product in $\mathbb{R}_\lambda^3$ becomes a block-diagonal infinite-matrix product
\beqa
{\phi_1\star \phi_2}\star...\star\phi_n(x)&=&\sum_{j,m_i, \tilde m_i}{ \phi_1}^{j}_{m_1\tilde m_1} {\phi_2}^{j}_{m_2\tilde m_2}...{\phi_n}^j_{m_n\tilde m_n}
v^{j}_{m_1\tilde m_1} \star v^{j}_{m_2\tilde m_2}\star...\star v^{j}_{m_n\tilde m_n} \; \nn\\
&=& {\sum_{j,m_1, \tilde m_n} (\Phi_1^j\cdot \Phi_2^j\cdot...\cdot\Phi_n^j)_{m_1 \tilde m_n} v^j_{m_1 \tilde m_n}} \label{starprodtr}
\eeqa
 where the infinite matrices $\Phi$ have been rearranged into a block-diagonal form, each block being the $(2j+1)\times (2j+1)$ matrix   {$\Phi^j=\{\phi^j_{mn}\}, \, -j\le m,n\le j$}.
  \subsection{Integration}
 The definition of the integral in the algebra $\gR^3_\lambda$ is the one introduced in \cite{V14} by one of the authors. It has been slightly modified with respect to our previous definition in \cite{vit-wal-12}  in order to better reproduce the commutative limit.  It is indeed a  generalization to the noncommutative case of the results contained in  \cite{DMV05}, where the  KS map is used. 
In \cite{DMV05}  it is   observed  that $\pi_{KS}$, defined in \eqn{KS}, defines a principal fibration $\gR^4-\{0\}\rightarrow \gR^3-\{0\}$ with structure group $U(1)$. The fibre is therefore compact. Moreover $\mathcal{F}(\gR^3-\{0\})$ can be mapped to  $\mathcal{F}(\gR^4-\{0\})$ by
 \be
\pi^*_{KS}: f \in \mathcal{F}(\gR^3-\{0\})\rightarrow f\circ \pi_{KS}\in \mathcal{F}(\gR^4-\{0\})
\ee
with  $\pi_{KS}^\star$ the pull-back map.
This realizes $\mathcal{F}(\gR^3-\{0\})$ as the subalgebra of $\mathcal{F}(\gR^4-\{0\})$ of functions which are constant along the fibers.
The vector field which generates the fiber $U(1)$
\be
Y_0= y^0 \frac{\del}{ \del y^3} - y^3 \frac{\del}{ \del y^0 } +y^1 \frac{\del}{ \del y^2} -y^2 \frac{\del}{ \del y^1} \label{fibergen}
\ee
defines indeed $\mathcal{F}(\gR^3-\{0\})$ as its kernel (it corresponds in the noncommutative case to the inner derivation $[x_0, \cdot]_\star$).
It is shown that, given the ordinary volume forms on $\gR^3$ and $\gR^4$, $\mu_3= dx^1\wedge dx^2\wedge dx^3$ and $\mu_4= dy^0\wedge dy^1\wedge dy^2\wedge dy^3 $, we have
\be
\pi^\star_{KS} (\mu_3 ) \wedge \alpha_0\propto  R^2 \mu_4
\ee
with $\alpha_0$ the dual form of the vector field $Y_0$ and volume form on the fiber. With our conventions the proportionality factor is $1/4$.
Observing that functions on $\gR^3$ are constant along the fiber $U(1)$, we can factorize the integral along the fiber which just gives a factor of $2\pi$, so that
 we have
 \be
\int \mu_3 \;f =\frac{1}{2\pi}\int \mu_4\; \pi^*_{KS}(x_0) \;  \pi^*_{KS}(f) \label{intdef}
\ee
where \eqn{rR} has been used.

Therefore, we may generalize to  the noncommutative case, assuming    \eqn{intdef}  as a definition,
\be
\int_{\gR^3_\lambda} f:=\frac{1}{2\pi}\int_{\gR^{4}_\theta } \pi^*_{KS}(x_0) \;  \bullet  \pi^*_{KS}(f)  \label{intnoncom}
\ee
compatible with the commutative limit. The symbol $\bullet$ indicates the two possible choices that we have in generalizing \eqn{intdef}, which correspond to  star-multiply  the weight coming from the integration measure, $\pi^*_{KS}(x_0)$,  with the integrand  $ \pi^*_{KS}(f)$ or to use the commutative product. Since $x_0$ star-commutes with all elements of the algebra, the two definitions only differ by a constant shift, as we shall see in a moment. Let us point out that \eqn{intnoncom} differs from the one in   \cite{vit-wal-12}, by  the factor  $\pi^*_{KS}(x_0)$.
Both definitions are legitimate, but the one proposed here has the advantage of reproducing the usual integral on $\gR^3$ once the commutative limit is performed.
Eq. \eqn{intnoncom} implies for the basis functions
\be
\int_{\Rl} v^j_{m\tilde m} (x)= \left\{
\begin{array}{ll}
8\pi \lambda^3 j\; \delta_{m \tilde m}   &{\rm a)}\\
8\pi\lambda^3 (j+1)\; \delta_{m \tilde m}& {\rm b)}
\end{array}
\right.
\ee
where the result a) corresponds to  the choice to star-multiply the weight-function $x_0$ with the integrand (it may be easily verified using the last of Eqs. \eqn{x0v}), whereas the result b) corresponds to choosing the point-wise multiplication and it is obtained by re-expressing the result of the product in terms of the basis elements. As announced, it amounts to a constant shift. We shall choose the second option in the paper.
We thus have
\be
\int_{\Rl} v^j_{m\tilde m}  \star v^j_{n\tilde n} (x)   =
8\pi\lambda^3(j+1)\delta_{\tilde m n }\delta_{m \tilde n} \label{properties2b}
\ee
Thanks to these results we obtain, for the integral of the star product \eqn{starprodtr}
\be
\int_{\mathbb{R}^3_\lambda} \phi_1\star \phi_2\star...\star\phi_n = 8\pi\lambda^3\sum_j (j+1) \tr_j (\Phi_1^j\cdot \Phi_2^j\cdot...\Phi_n^j) \label{integralofprod}
\ee
with $\tr_j$ the trace in the $(2j+1)\times(2j+1)$  subspace.
Notice that, on performing the sum up to $j=N/2$, with $\Phi^j=\Psi^j=  \bf{1}_j$ we obtain the result $8\pi\lambda^3\sum_{n=0}^N(n/2+1)(n+1)\simeq\frac{4}{3}\pi\lambda^3 N^3$, which reproduces correctly  the volume of a sphere of radius $\lambda N$.
\subsection{Derivations} In the commutative case derivations of the algebra $\mathcal{F}(\gR^3-\{0\})$ are obtained by projecting the derivations of $\mathcal{F}(\gR^4)$ through the KS map \eqn{KS}. It may be seen \cite{DMV05} that projectable vector fields are defined by the condition [$D_i, Y_0]=0$, with $Y_0$ given in \eqn{fibergen} . They correspond to the three rotations generators $Y_i$,  and the dilation $D$
\be
\pi_{KS*} (Y_i)= X_i=  \epsilon_{ij}^k x^j \frac{\del}{\del x^k}\, , \;\;\;\;
\pi_{KS*} (D)= x^i\frac{\del}{\del x^i} \label{dilation}
\ee
with $\pi_{KS*} $ the push-forward map, $Y_i =y^0 \frac{\del}{\del y^i}  -y^i \frac{\del}{\del y^0} -\epsilon_{ij}^k y^j  \frac{\del}{\del y^k}$ and $D= y^\mu\frac{\del}{\del y^\mu} $. As well known, the three rotations are not independent since $x^i \cdot X_i =0$. When passing to the noncommutative case the three rotations  are still derivations of the algebra $\gR^3_\lambda$ and may be given the form of inner derivations, but the dilation, in the form of \eqn{dilation}, is not anymore a derivation.
On using the star product \eqn{starsu2} we have for rotations
\be
X_i (\phi)= -\frac{i}{\lambda}[x_i,\phi]_\star\, ,  \,\,\; i=1,..,3 \label{rots}
\ee
which obviously satisfies the Leibnitz rule. Moreover they {\it become independent} (even though $x^i \star X_i (\phi)+   X_i (\phi)\star x^i=0$, derivations are not a module over the algebra in the NC case, they are only a left module over the center of the algebra). As for the dilation, it is easy to check that it does not satisfy the Leibnitz rule (for example, on applying it to the star product of coordinates). Therefore, the only derivations of the algebra $\Rl$ closing a Lie algebra are the three inner derivations $X_i$. Let us notice that they are also {\it sufficient} in the sense that only functions which are in the center of $\Rl$ are simultaneously annihilated by all of them.  As clarified in appendix \ref{appendix0}, the notion of sufficiency  replaces the notion of basis of a module in the noncomutative setting. 

Let us notice that there is a way to implement the dilation as a derivation of the star product \eqn{starsu2}. This amounts to enlarge the algebra $\gR^3_\lambda$ to include the noncommutativity parameter $\lambda$ (see \cite{GLRV06} where the construction is performed for the Moyal algebra $\gR^4_\theta$). It may be checked that, in such a case, the vector field
\be
D_\lambda= x_i\frac{\del}{\del x_i} + \lambda \frac{\del}{\del \lambda}\equiv x_0\frac{\del}{\del x_0} + \lambda \frac{\del}{\del \lambda}
\ee
is an outer derivation of the enlarged algebra (see \cite{V14} for further details). Moreover, together with the three derivations $X_i$, it closes the standard $\mathfrak{u}(2)$ Lie algebra, as in the commutative case. The inclusion of such a derivation with a suitable modification of the definition of the algebra $\Rl$ shall be considered elsewhere. 

Expressing the fields of NCFT on $\Rl$ in the canonical basis yields diagonal interaction vertices so that this latter may be physically viewed as the interaction basis.
We will also use  another basis, the fuzzy spherical harmonics, $Y^j_{lk}$, widely used in the literature related to the fuzzy sphere (see  appendix \ref{appendix0} for details).  It turns  out that a class of natural Laplacian operators on $\mathbb{R}^3_\lambda$, such as the one considered in \cite{vit-wal-12} is diagonal in such a  basis, together with the  gauge-fixed kinetic operator of the gauge models built in this paper. This basis may then be viewed physically as the propagation basis. 

The issue of the definition of a Laplacian for the algebra $\Rl$ is an important one. It has been already addressed in \cite{vit-wal-12} and recently reconsidered in \cite{V14}. The problem is that,  on one hand we would like a Laplacian which gives back the ordinary Laplacian on $\gR^3$ when the commutative limit is performed. On the other hand, one would like to construct a Laplacian in terms of the derivations of the algebra (see however \cite{pres} where a different proposal not based on derivations is explored). We have seen that the algebra has only inner derivations, which, in the commutative limit, reproduce rotations. Hence, a Laplacian constructed in terms of them will not reproduce the radial part of the Laplacian in the commutative limit. It was thus argued in \cite{vit-wal-12} and further clarified in \cite{V14} that a multiplicative operator quadratic in $x^0$ should be added, because the star-product of $x^0$ with elements of the algebra contains the dilation operator. The issue of the commutative limit is however still to be understood.  In any case, we shall see in next sections that the effective action we are going to consider for the fluctuations of the gauge fields only contains the natural, derivations based,  Laplacian. It is still to be understood how to implement the modification proposed in \cite{vit-wal-12} or \cite{V14} at the level of gauge theory. 
\par

%In the subsequent analysis, it will be convenient to use the equivalent description of $\Rl$ as $\mathcal{W}_V(\Rl)$, i.e already represented on the operators of ${\cal{H}}$ via the map $\mathcal{W}_V$. Any element of $%\mathcal{W}_V(\Rl)$ is then given by the rightmost equality of \eqref{phixi} and the $\star$-product reduces to the matrix product \eqref{starmatrix}. 
%To simplify the notations, we set from now on $\mathcal{W}_V(\Rl)=\Rl$, %and drop the hat and $\star$ symbols. 
%To simplify the notations, we drop from now on the $\star$ product, whenever no ambiguity is possible. 

\section{\texorpdfstring{Classical gauge-invariant models}{Diff calcul next}}\label{section3}
\subsection{Connection and curvature}\label{subsection231}
We consider now the natural differential calculus generated by the Lie algebra of real inner derivations of $\Rl$ defined in terms of $X_i$ \eqn{rots} by
\begin{equation}
\DER(\Rl):=\{D_i:=-\frac{1}{\lambda}X_i =\frac{i}{\lambda^2}[x_j,\cdot]_\star,\ j=1,2,3\}\label{derivatives}
\end{equation}
with the  relation
\begin{equation}
[D_i,D_j]=-\frac{1}{\lambda}\epsilon_{ij}^kD_k\;\;\; \forall\, i,j,k=1,2,3\label{commutderiv}, 
\end{equation}
Notice that $D_0= \frac{i}{\lambda^2}[x_0, \cdot]_\star$ is a trivial derivation of the algebra, being 
${[}D_i, D_0{]}= 0 \;\;\forall\, i=1,2,3$ and $D_0 (f)=0 \;\; \forall f\in \Rl$.  

 The mass dimensions are $[\lambda]=-1$ and $[D_i]=1$. It is straightforward to check that $\DER(\Rl)$ is a module over ${\cal{Z}}(\Rl)$, the center of 
$\Rl$, which is generated by the element $x_0$ introduced above. Therefore, as stated above, the three derivations $D_i$ are independent as a module over the center. Moreover, they are also {\it sufficient}, that is they verify $D_i (f)= 0\, \forall i=1,..,3$ if and only if $f\in {\cal{Z}}(\Rl)$.   From this follows that $\DER(\Rl)$ of Eq. \eqref{derivatives} generates a differential calculus as described above. 

In order to generalize $U(1)$ gauge theories, as a right module on $\Rl$, we pick the algebra $\Rl$ itself. Then, from the above scheme, one easily checks from the first of Eqs. \eqref{connect} that a Hermitian connection on $\Rl$ for the Hermitian structure $h(a_1,a_2)=a_1^\dag\star  a_2$, $\forall a_1,a_2\in\Rl$ is entirely determined by the elements $A_i:=\nabla_{D_i}(\bbbone)$ with
\begin{equation}
\nabla_{D_i}(a):=\nabla_i(a)=D_i a+A_i\star  a,\ A_i^\dag=-A_i\label{connectpartic},
\end{equation}
while from Eq. \eqref{etaX} one infers that 
\begin{equation}
\eta(D_i):=\eta_i=\frac{i}{\lambda^2}x_i\label{def-eta}
\end{equation}
and the invariant connection and covariant one form are respectively
\begin{eqnarray}
\nabla^{inv}_{i}(a)&=&D_i a-\eta_i \star a=-\frac{i}{\lambda^2}a\star x_i;\\
{\cal{A}}_i&:=&(\nabla_{i}-\nabla^{inv}_{i})(a)=A_i+\frac{i}{\lambda^2}x_i\label{covarcoord},
\end{eqnarray}
so that
\begin{equation}
\nabla_i(a)={\cal{A}}_i \star a-\frac{i}{\lambda^2}a\star x_i
\end{equation}
for any $a\in\Rl$. By noting that $[\eta_i,\eta_j]_\star=\eta_{[D_i,D_j]}=-\frac{1}{\lambda}\epsilon_{ijk}\eta_k$, we find from \eqref{finv} that
\begin{equation}
F^{inv}_{ij}=[\eta_i,\eta_j]_\star-\eta_{[D_i,D_j]}=0\label{invcurvat}.
\end{equation}
Finally, by combining \eqref{courgene} with
\begin{eqnarray}
[\nabla_i,\nabla_j]a&=&-\frac{1}{\lambda}\epsilon_{ijk}D_k a+[A_i,A_j]_\star \star a+(D_i A_j-D_j A_i)\star a\label{interm1}\\ \nabla_{[D_i,D_j]}a&=&-\frac{1}{\lambda}\epsilon_{ijk}(D_k a+A_k \star a)\label{interm2},
\end{eqnarray}
we obtain the expression for the curvature
\begin{eqnarray}
F_{ij}&=&(D_i A_j-D_j A_i)+[A_i,A_j]_\star+\frac{1}{\lambda}\epsilon_{ijk}A_k\label{curvatur-A}\\
&=&[{\cal{A}}_i,{\cal{A}}_j]+\frac{1}{\lambda}\epsilon_{ijk}{\cal{A}}_k\label{curvatr3l}.
\end{eqnarray}
The gauge transformations are still given by \eqref{gaugeconnect}, \eqref{gaugecurv} with $g\in{\cal{U}}(\Rl)$, i.e $g^\dag g=gg^\dag=\bbbone$. We finally make the rescaling $A_i\to -iA_i$ so that now $A_i^\dag=A_i$ and the curvature \eqref{curvatur-A} becomes 
\begin{equation}
F_{ij}=-i(D_i A_j-D_j A_i)-[A_i,A_j]_\star-i\frac{1}{\lambda}\epsilon_{ijk}A_k\label{praticurvat}.
\end{equation}

\subsection{A family of gauge-invariant actions}\label{subsection232}

We now look for families of gauge-invariant functional actions 
 depending on $A_i$, $S_{cl}(A_i)$, i.e we assume that $A_i$ is the relevant field variable. We do not adopt here the viewpoint developed in \cite{MVW13} leading to a matrix model formulation of gauge theories on $\mathbb{R}^2_\theta$ with ${\cal{A}}_i$ chosen as the field variable. Our principal requirements are:\\
i) The gauge invariant functional actions 
%built from the available structures $\delta_{ij}$ and $\varepsilon_{ijk}$ 
are at most quartic in $A_i$;\\
ii) No linear terms in $A_i$ are involved;\\
iii) The kinetic operator is positive (upon gauge fixing).\par 

The requirement i) is obviously fulfilled by any functional of the form $\int_{\Rl}[{\cal{P}}({\cal{A}}(A))]$, where ${\cal{P}}({\cal{A}})$ is a star-polynomial with degree $\delta=4$.\par 
The requirement ii) insures that the classical equations of motion support the solution $A^0_i=0$, which otherwise would imply a non trivial vacuum for the classical action. Then, one would have to expand the classical action $S_{cl}(A_i)$ around $A^0_i$, i.e setting $A_i=A^0_i+\alpha_i$ where now $\alpha_i$ may be interpreted as covariant coordinates (being related to the difference of two connections). This does not fit with our field variable assumption. Recall that in the case of Moyal space $\mathbb{R}^4_\theta$, non trivial vacuum solutions are known to occur within gauge theories \cite{Wallet:2007c}, \cite{GWW2} generating huge difficulties. Recall also that for the commutative Yang-Mills action, a salient property valid at the quantum level is that the tadpole (1-point) function stemming from the cubic gauge fields coupling vanishes automatically thanks to the Lie algebraic structure of the interaction vertex. In the present situation, the structure of this latter is quite different. Since we examine the possibility to have natural noncommutative analogs of Yang-Mills theory on $\mathbb{R}^3_\lambda$, an important issue to examine is the fate of the tadpole which actually takes part to the quantum stability of the vacuum. \par 

Another point to examine is the possibility to build a massless theory as commutative Yang-Mills is. In fact, the computation of loop diagrams is complicated, even for the 1-point function. This is due to the structure of the vertices and the kinetic operator. Fortunately, one useful simplification occurs when no mass term is present in the classical theory, so that the zero mass issue can be examined. \par

The above discussion points towards the following gauge invariant functional {\footnote{Summation over repeated indices is understood. Moreover, we shall omit to indicate the star product from now on, unless required to avoid ambiguities.}} satisfying requirement i):
\begin{equation}
S_{cl}(A_i)=\frac{1}{g^2}\int_{\Rl} \big(\alpha{\cal{A}}_i{\cal{A}}_j {\cal{A}}_j {\cal{A}}_i+\beta{\cal{A}}_i {\cal{A}}_j {\cal{A}}_i {\cal{A}}_j+\zeta\varepsilon_{ijk}{\cal{A}}_i {\cal{A}}_j {\cal{A}}_k+m{\cal{A}}_i {\cal{A}}_i     \big)\label{spectral-action0}
\end{equation}
where
${\cal{A}}_i$ must be viewed as a functional of $A_i$, ${\cal{A}}_i=-iA_i+\eta_i$ and $\alpha$, $\beta$, $\zeta$, $m$ are real parameters with respective mass dimensions $[\alpha]=[\beta]=0$, $[\zeta]=1$, $[m]=2$. The overall constant $1/g^2$ has mass dimension -1 and it is necessary in $D=3$ in order to get a dimensionless action.  
On using the matrix basis introduced previously and on expanding the gauge fields in such a basis
\beqa
A_i&=&\sum_{j\in\frac{\mathbb{N}}{2}}\sum_{-j\le m,\tilde m\le j}(A^j_i)_{m\tilde m}v^j_{m\tilde m} \label{genr-expans}\\
\eta_i&=&\sum_{j\in\frac{\mathbb{N}}{2}}\sum_{-j\le m,\tilde m\le j}(\eta^j_i)_{m\tilde m}v^j_{m\tilde m}
\eeqa
 the integral may be reduced to a sum of traces, by means of Eq. \eqn{integralofprod}
\be
S_{cl}(A_i)=\widetilde\tr \big(\alpha{{\cal A}^j}_i{{\cal A}^j}_k {{\cal A}^j}_k {{\cal A}^j}_i+\beta{{\cal A}^j}_i {{\cal A}^j}_k {{\cal A}^j}_i {{\cal A}^j}_k+\zeta\varepsilon_{ilk}{{\cal A}^j}_i {{\cal A}^j}_l {{\cal A}^j}_k+m{{\cal A}^j}_i {{\cal A}^j}_i     \big)\label{spectral-action}
\ee
where  we have introduced the shorthand for the weighted trace
\be
\widetilde\tr=\frac{8\pi\lambda^3}{g^2}\sum_j(j+1)\tr_j .
\ee
We recall that  $A_i$,  $\eta_i$, ${\cal A}_i=iA_i + \eta_i$,  are infinite-dimensional, block diagonal matrices, each block  $A_i^j, \eta^j_i,  {\cal A}_i^j$ being a $2j+1\times 2j+1$ matrix. To simplify the notation, we shall omit the superscript $j$ from now on unless otherwise stated. 
The terms linear in $A_i$ are given by
\begin{equation}
S^1_{cl}(A_i)=i\widetilde\tr\big(-4(\alpha+\beta)(\eta^2)\eta_i A_i+(3\frac{\zeta}{\lambda}-4\frac{\beta}{\lambda^2}-2m)\eta_i A_i \big)\label{linear-generalaction}.
\end{equation}
The requirement ii) is fulfilled provided
\begin{eqnarray}
\alpha+\beta&=&0\label{condition1},\\
3\frac{\zeta}{\lambda}-4\frac{\beta}{\lambda^2}&=&2m\label{condition2}.
\end{eqnarray}
Condition \eqref{condition1} is automatically satisfied whenever the quartic part of the action \eqref{spectral-action} comes  from $\widetilde\tr(F^\dag_{ij}F_{ij})$. That will be assumed from now on{\footnote{This term is {\it{formally}} similar to a Yang-Mills action, up to the last term in \eqref{praticurvat}.}}. We will also assume $\alpha=-\beta=2$. Then, setting for convenience 
\begin{equation}
\gamma:=\zeta+\frac{4}{\lambda},\ \mu:=m+\frac{2}{\lambda^2}\label{parameter-natural},
\end{equation}
we obtain from \eqref{spectral-action} the following gauge-invariant action
\begin{equation}
S_{cl}(A_i)=\widetilde\tr\big(F^\dag_{ij}F_{ij}+\gamma\epsilon_{ijk} {\cal{A}}_i{\cal{A}}_j{\cal{A}}_k+\mu{\cal{A}}_i{\cal{A}}_i\big)\label{actionansatz},
\end{equation}
which satisfies the requirement ii) provided
\begin{equation}
\mu=\frac{3}{2\lambda}\gamma\label{condition2bis}.
\end{equation}
One then obtains
\begin{equation}
S_{cl}(A_i)=\widetilde\tr\big(F^\dag_{ij}F_{ij}+\gamma(\epsilon_{ijk} {\cal{A}}_i{\cal{A}}_j{\cal{A}}_k+\frac{3}{2 \lambda}{\cal{A}}_i{\cal{A}}_i)\big)\label{actionansatz-bis}.
\end{equation}
The kinetic term of \eqref{actionansatz-bis} is given by
\begin{equation}
S^2_{cl}(A_i)=\widetilde\tr\big(A_i[-2\delta_{ij}D^2+(\frac{2}{\lambda^2}-\mu)\delta_{ij}+2D_i D_j
-{\lambda}(\frac{2}{\lambda^2}-\mu)\varepsilon_{ijk}D_k]A_j\big)\label{kinetic-general},
\end{equation}
which involves a mass-type term $\sim(\frac{2}{\lambda^2}-\mu)A_i A_i$.\par 
At this stage, one interesting remark is in order. We note that the total action 
\begin{equation}
S_{cl}(A_i)=\widetilde\tr F^\dag_{ij}F_{ij}+S^{CS}_{cl}(A_i)
\end{equation}
is (up to unessential changes in the parameters) very similar to the Alekseev-Recknagel-Schomerus gauge action on the fuzzy sphere \cite{ARS00} whenever we retain only the projection of the gauge fields on one single fuzzy sphere of the ''foliation''\footnote{We are grateful to Harold Steinacker for bringing to our attention   this important point.}. This amounts to fix the radius eigenvalue $j$ in  the field expansion \eqn{genr-expans},.
Such a model has been widely studied (see for example \cite{CDY04},  and references therein). The comparison with our results is certainly to be done, it is however not straightforward for many reasons:  we perform our calculation in the interaction basis whereas all other calculations available in the literature are performed in the propagation basis (fuzzy harmonics),  also, with respect to \cite{CDY04}  a different gauge choice has been made. We therefore postpone the analysis to a subsequent work.\par 

In order to prepare the ensuing discussion, it is interesting to consider the case obtained by dropping the term $\sim\widetilde\tr(F^\dag_{ij}F_{ij})$;  what is left is a formal analog of a Chern-Simons term. This amounts to set $\alpha=\beta=0$ in \eqref{spectral-action}. One easily obtains the corresponding classical action fulfilling requirement ii):
\begin{equation}
S^{CS}_{cl}(A_i)=\frac{3\gamma}{2}\widetilde\tr\big(A_i[-\frac{1}{\lambda}\delta_{ij}+\varepsilon_{ijk}D_k ]A_j \big)+i\gamma\widetilde\tr(\varepsilon_{ijk}A_i A_j A_k)\label{chern-clasq-final}.
\end{equation}

The action \eqref{chern-clasq-final} must be supplemented by a BRST invariant gauge-fixing term. Here, it is especially convenient to choose an axial-type gauge, namely $A_3=0$. We follow here the usual liturgy for the BRST gauge-fixing in NCFT. Rather universal algebraic tools in BRST symmetry can be found in \cite{stor-wal}. The gauge-fixed action is written as
\begin{equation}
S^{CS}_{tot}=S^{CS}_{cl}+s\widetilde\tr\big(\bar{C}A_3\big)=S^{CS}_{cl}+\widetilde\tr\big( bA_3-\bar{C}D_3C+i\bar{C}[A_3,C] \big)\label{gaugefixed2}
\end{equation}
where $s$ is a nilpotent Slavnov operation{\footnote{Recall that $s$ acts as a graded derivation with respect to the grading defined by the sum of the degree of forms and ghost number (modulo 2).  }} with structure equations defining the BRST symmetry given by
\begin{equation}
sA_i=D_i C-i[A_i,C]=[{\cal{A}}_i,C],\ sC=iCC,\ s\bar{C}=b,\ sb=0, \label{brst}
\end{equation}
in which $\bar{C}$, $C$ and $b$ are respectively the antighost, ghost and St\"uckelberg field with respective ghost numbers $-1$, $1$ and $0$. Then, by combining \eqref{gaugefixed2} with \eqref{brst} and formally integrating over the St\"uckelberg field $b$ which amounts to set $A_3=0$ into the action \eqref{gaugefixed2}, it is easy to realize that the interaction term vanishes 
while the ghost part decouples from the gauge potential part leading to a gauge-fixed free theory. Note that it is somewhat similar to what happens for the (commutative) non-Abelian Chern-Simons theory on $\mathbb{R}^3$ (see for instance \cite{albe-sengup} and references therein).\par 

Consider now only the Yang-Mills type term in \eqref{actionansatz} ($\gamma=\mu=0$). Notice that it corresponds to a massive theory, with mass term $\sim\frac{2}{\lambda^2}\widetilde\tr(A_i A_i)$. The corresponding kinetic operator in the axial gauge given by 
$$
K^{YM}_{ij}=(-2\delta_{ij}(D^2-\frac{1}{\lambda^2})-\frac{2}{\lambda}\varepsilon_{ij}D_3 )+2D_iD_j,   ~~~~i,j=1,2
$$ 
is very hard to invert,  the difficulty coming  essentially from the term $D_iD_j$. Working in a Landau-type gauge as the one used below, permits to get rid of the terms $\sim D_i D_j$ and the corresponding gauge fixed kinetic operator becomes diagonal in the space indices, except terms ``linear in the derivative'' $\sim\varepsilon_{ijk}D_k$ This again makes the computation of the propagator very difficult. Note that these linear terms reflect the non-commutativity of the derivatives, which is one source of the technical difficulties. 

In the present case, the use of the matrix bases introduced in the subsection \ref{subsection22} yields kinetic operators that do not obey the indices conservation law (see e.g \eqref{consindice1}, \eqref{consindice2} below) but shifted conservation laws. Hence, they cannot be related to Jacobi operators unlike the kinetic operator of the Grosse-Wulkenhaar model  or in \cite{vit-wal-12}, \cite{MVW13}. They are instead related to a  kind of generalized Jacobi operators. Notice that a similar feature appears within the matrix model formulation of gauge theories on $\mathbb{R}^2_\theta$ developed in \cite{MVW13} for an interesting class of vacua. This deserves further investigations.\par

An interesting simplification arises when the action \eqref{actionansatz-bis} is formally massless. In view of \eqref{kinetic-general}, this occurs whenever 
\begin{equation}
\mu=\frac{2}{\lambda^2}\label{conditionmassless},
\end{equation}
for which terms linear in the derivative also disappear, thus   simplifying  the computation of the propagator for $A_i$. From now on, we will assume that \eqref{conditionmassless} holds true, therefore focusing on a gauge-invariant massless  theory on $\mathbb{R}^3_\lambda$.\par 

It is convenient to use a Feynman-Landau type gauge $D_i A_i=0$. Then the corresponding BRST invariant gauge-fixed action is
\begin{eqnarray}
S_{tot}&=& S_{cl}(A_i)+s\widetilde\tr\big(\bar{C}(D_i A_i)+\xi \bar{C}b \big)
=\widetilde\tr\big(A_i[-2\delta_{ij}D^2+(2+\frac{1}{4\xi})D_i D_j]A_j\nonumber\\
&+&4i\, D_i A_j[A_i,A_j] -i\frac{4}{3\lambda}\varepsilon_{ijk}A_i A_j A_k\nonumber\\
&-&2(A_i A_i)^2+2A_i A_j A_i A_j \big)+\widetilde\tr\big(-\bar{C}(D^2)C+i\bar{C}D_i[A_i,C] \big)\label{gaugefixfinal}.
\end{eqnarray}
where $\xi$ is a dimensionless gauge parameter, the BRST symmetry is still defined by \eqref{brst} and integration over the $b$ field has been performed as above. Unessential terms $\sim \widetilde\tr(\varepsilon_{ijk} x_i x_j x_k)$ have been dropped. Now, one easily observes that the kinetic operator becomes diagonal in the ``space indices'' when
\begin{equation}
\xi=-\frac{1}{8}\label{special-gauge},
\end{equation}
which hereafter is referred as the diagonal gauge. The inversion of this operator, that can be related to an operator of Jacobi type, becomes now possible. Notice that in the present massless situation, the occurrence of IR singularity can be expected in the propagator as shown in a while. In fact, this IR singularity will be harmless in the ensuing analysis.\par

\subsection{Gauge and ghost propagators}\label{subsection33}

The propagator for the $A_i$ can be computed by expressing in the canonical matrix basis for $\Rl$ the corresponding quadratic part of the gauge-fixed action. By observing that  $
D^2=-\Delta$ where $\Delta$ is the Laplacian operator already considered  in \cite{vit-wal-12} for a suitable choice of parameters, 
\begin{equation}
\Delta v^j_{m\tilde m}= \frac{1}{\lambda^4} \Bigl( \frac{1}{2}( [x_+,[x_-, v^j_{m\tilde m}]_\star]_\star + [x_-,[x_+, v^j_{m\tilde m}]_\star]_\star )+ x_3,[x_3, v^j_{m\tilde m}]_\star]_\star \Bigr)
\end{equation}
and further using the expansion of the field variable $A_i$ in the canonical basis \eqn{genr-expans},
the kinetic part of the action \eqref{gaugefixfinal} in the diagonal gauge becomes
\begin{equation}
S_2(A_i)=2\sum_{j_1,j_2}\sum_{m_1,\tilde m_1}\sum_{m_2,\tilde m_2}(A_i)^{j_1}_{m_1\tilde m_1}(\Delta)^{j_1,j_2}_{m_1\tilde m_1;m_2\tilde m_2}    (A_i)^{j_2}_{m_2 \tilde m_2}\label{quadrat-matrix},
\end{equation}
with (see Eqs. \eqref{x0v})
\begin{eqnarray}
(\Delta)^{j_1,j_2}_{m_1\tilde m_1;m_2\tilde m_2}&=&\delta_{j_1j_2}\big(\delta_{\tilde m_1m_2}\delta_{m_1\tilde m_2}D^{j_2}_{m_2\tilde m_2}\nonumber\\
&-&\delta_{\tilde m_1,m_2+1}\delta_{m_1,\tilde m_2+1}B^{j_2}_{m_2\tilde m_2}-\delta_{\tilde m_1,m_2-1}\delta_{m_1,\tilde m_2-1}H^{j_2}_{m_2\tilde m_2}\big),\\
D^{j}_{m_2\tilde m_2}&=& \frac{8\pi\lambda}{g^2}(j+1) \Bigl(2j^2+2(j-m_2\tilde m_2) \Bigr)\\
B^{j}_{m_2\tilde m_2}&=&\frac{8\pi\lambda}{g^2}(j+1) \big[(j+m_2+1)(j-m_2)(j+\tilde m_2+1)(j-\tilde m_2) \big]^{\frac{1}{2}} \\
H^{j}_{m_2\tilde m_2}&=&\frac{8\pi\lambda}{g^2}(j+1) \big[(j+m_2)(j-m_2+1)(j+\tilde m_2)(j-\tilde m_2+1) \big]^{\frac{1}{2}}.
\end{eqnarray}
The indices conservation law \cite{vit-wal-12} is
\begin{equation}
\Delta^{j_1 j_2}_{mn;kl}\ne 0 \implies
    j_1=j_2,\;\;\;  m+k= n+l\label{consindice1}.
\end{equation}
Another useful relation is
\begin{equation}
\Delta^{j_1 j_2}_{mn;kl}=\Delta^{j_1 j_2}_{kl;mn}\label{cons-indiceprim}.
\end{equation}
The gauge fixed kinetic operator is a positive operator, in agreement with the requirement iii) (see beginning of subsection \ref{subsection232}). In particular, one has (see Eq. \eqref{xxharm})
\begin{equation}
\Delta Y^j_{lk}= \frac{1}{\lambda^2} l(l+1) Y^j_{lk}\    \;\;\; j\in{{\mathbb{N} }\over{2}},\ 0\le l\le 2j,\ l\in\mathbb{N},\  -l\le k\le l  \label{finallaplacescale}
\end{equation}
so that the spectrum of $\Delta$ is positive, $spec(\Delta)\subset\mathbb{R}^+$. Notice that it involves the zero eigenvalue, as it can be expected in a massless theory. Combining the mass dimension of $\lambda$ with \eqref{finallaplacescale} singles out a natural choice for the UV and IR regions, corresponding respectively to large and small indices $l$.\par 

In the same way, the kinetic part for the ghost sector can be expressed  as
\begin{equation}
S_2(\bar C,C)=\widetilde\tr(\bar C\Delta C)=\sum_{j_1,j_2}\sum_{m_1,\tilde m_1}\sum_{m_2,\tilde m_2}\bar C^{j_1}_{m_1\tilde m_1}(\Delta)^{j_1,j_2}_{m_1\tilde m_1;m_2\tilde m_2}  C^{j_2}_{m_2 \tilde m_2}\label{quadrat-ghost}
\end{equation}
where the Grassman variables $\bar C^{j_1}_{m_1\tilde m_1}$ and $C^{j_2}_{m_2 \tilde m_2}$ inherit the respective ghost number of $\bar C$ and $C$.\par 

The propagator $P^{j_1 j_2}_{mn;kl}$ is defined as the inverse of $\Delta^{j_1 j_2}_{mn;kl}$ by the relations
\begin{equation}
\sum_{k,l=-j_2}^{j_2}\Delta^{j_1 j_2}_{mn;lk}P^{j_2 j_3}_{kl;rs}=\delta^{j_1 j_3}\delta_{ms}\delta_{nr},\ \sum_{m,n=-j_2}^{j_2}P^{j_1 j_2}_{rs;mn}\Delta^{j_2 j_3}_{nm;kl}=\delta^{j_1 j_3}\delta_{rl}\delta_{sk}\label{definvers},
\end{equation}
from  which follows
\begin{equation}
P^{j_1 j_2}_{mn;kl}\ne0\implies     j_1=j_2,\;\;\;m+k= n+l\label{consindice2}.
\end{equation}
The explicit expression for the propagator can be readily obtained by an  adaptation of the results obtained in \cite{vit-wal-12} (note the factor $(j_1+1)$ in the denominator due to the weighted trace we have introduced). It can be written as
\begin{equation}
P^{j_1 j_2}_{mn;pq}= \frac{g^2}{8\pi\lambda}\delta^{j_1 j_2}\frac{1}{(j_1+1)(2j_1+1)}\sum_{l=0}^{2j_1} \sum_{k=-l}^l\frac{1}{l(l+1)} (Y^{j_1\dag}  _{l k})_{nm} (Y^{j_2}_{l k})_{qp}\label{propagator}.
\end{equation}
Note that Eq.  \eqref{propagator} is singular when $l=0$ which corresponds to an IR singularity, as it can be expected in a massless (gauge) theory. In the ensuing analysis, it will be understood that an IR regulator is used whenever \eqref{propagator} and/or sums $\sim\sum_{l=0}^{2j}$ are explicitly written. In fact, IR singularities will disappear as we will show in a while.\par

\section{Tadpole function at the one-loop order}\label{section4}

We now use the perturbative framework detailed  in the appendix \ref{appendix1} to compute the tadpole one-point function for the gauge potential. It receives contribution from ghost and gauge potential loops corresponding respectively to the ghost-gauge vertex and the trilinear vertex in the gauge-fixed action \eqref{gaugefixfinal}. Recall that we have chosen the diagonal gauge, defined by \eqref{gaugefixfinal} together with \eqref{special-gauge}. The computation is a bit cumbersome as the tadpole contributions for each $A_i$ must be considered separately. Nevertheless, a simplification occurs once it is noticed that the totally antisymmetric part of the cubic gauge potential vertex does not contribute.\par 
It is easier to begin with the computation of the tadpole with external $A_3$. We first consider the corresponding ghost loop contribution. By using the properties of the canonical basis recalled in the subsection \ref{subsection22} and in appendix \ref{appendix0}, together with the definition \eqref{derivatives} and observing that  \eqref{x0v}
\begin{equation}
[x_3,v^j_{mn}]=\lambda(m-n)v^j_{mn},\ \forall j\in\frac{\mathbb{N}}{2},\ -j\le m,n\le j
\end{equation}
the relevant ghost-gauge part of the action \eqref{gaugefixfinal} can be written as
\begin{equation}
S_{int}^{A_3\phi\pi}=\frac{k}{\lambda}\sum_{j,m,n,p}(j+1) (m-n)\bar C^j_{mn}((A_3)^j_{np}C^j_{pm}-C^j_{np}(A_3)^j_{pm})\label{sintghost}
\end{equation}
where the superscript $``\phi\pi''$ stands for ``Faddeev-Popov'' and $k=\frac{8\pi\lambda^3}{g^2}$.
Therefore,
\begin{equation}
S_{int}^{A_3\phi\pi}(\frac{\delta}{\delta \widetilde J},\frac{\delta}{\delta \bar{\widetilde \eta}},\frac{\delta}{\delta\widetilde\eta})=\frac{k}{\lambda}\sum_{j,m,n,p}(j+1) (n-m)(\frac{\delta}{\delta\bar{\widetilde \eta}^j_{mp}}\frac{\delta}{\delta (\widetilde J_3)^j_{pn}}\frac{\delta}{\delta\bar {\widetilde \eta}^j_{nm}}-\frac{\delta}{\delta\bar{\widetilde \eta}^j_{pn}}\frac{\delta}{\delta\bar {\widetilde\eta}^j_{nm}}\frac{\delta}{\delta (\widetilde J_3)^j_{mp}})\label{trilin-ghost-funct}
\end{equation}
where the source fields $J, \bar \eta, \eta$ have been rescaled by a factor of $k (j+1)$, as explained in appendix \ref{appendix1}.
From \eqref{trilin-ghost-funct} and \eqref{1loopgeneric}, one infers that the relevant part of $W(J,\eta,\bar\eta)$ related to the 1-loop ghost contribution to the above  1-point function is
\begin{equation}
W^{\phi\pi}_1((J_3))=-\frac{k}{4\lambda}\sum_{j,m,n,p,k,l}(j+1)(n-m)(P^{j,j}_{mp;nm}P^{j,j}_{pn;kl}-P^{j,j}_{pn;nm}P^{j,j}_{mp;kl})(\widetilde J_3)^j_{kl}\label{w1-ghost}.
\end{equation}
By Legendre transform and using \eqref{inverslegendre}, \eqref{legendreinter}, the part of the effective action defining the ghost contribution of the 1-point function is
\begin{eqnarray}
\Gamma^{1;\phi\pi}(A^3):&=&\sum_{n,p}\sigma^{j;\phi\pi}_{3\ np}(A_3)^j_{np}=\frac{k}{\lambda} \sum_{m,n,p} (j+1)(p-m)P^{j,j}_{mp;nm}(A_3)^j_{np}\nonumber\\
\sigma^{j;\phi\pi}_{3\ np}&=&\frac{k}{\lambda} \sum_{-j\le m\le j}(j+1)(p-m)P^{j,j}_{mp;nm}
\end{eqnarray}
where the last relation makes apparent the external indices $j,n,p$.\par 
Next, the relevant part of the cubic gauge potential interaction contributing to the 1-point function is
\begin{equation}
S_{int}^{AAA}=\frac{4 k}{\lambda}\sum_{i=1,2}\sum_{j,m,n,p}(j+1)(m-n)((A_i)^j_{mn}((A_3)^j_{np}(A_i)^j_{pm}-(A_i)^j_{np}(A_3)^j_{pm})\label{sintaaa},
\end{equation}
for which each of the two terms in the sum over $i$ will contribute equally to the 1-point function. As expected, the structure of the trilinear gauge potential coupling \eqref{sintaaa} is similar to ghost-gauge potential interaction \eqref{sintghost}, up to the Grassman nature of the ghost variables. Eq. \eqref{sintaaa} yields
\begin{equation}
S_{int}^{AAA}(\frac{\delta}{\delta \widetilde J})=\frac{4 k}{\lambda}\sum_{i=1,2}\sum_{j,m,n,p}(j+1) (m-n)(\frac{\delta}{\delta(\widetilde J_i)^j_{mp}}\frac{\delta}{\delta(\widetilde J_3)^j_{pn}}\frac{\delta}{\delta(\widetilde J_i)^j_{nm}}
-\frac{\delta}{\delta(\widetilde J_i)^j_{pn}}\frac{\delta}{\delta(\widetilde J_i)^j_{nm}}\frac{\delta}{\delta(\widetilde J_3)^j_{mp}})\label{trilin-aaa-funct}.
\end{equation}
By merely comparing \eqref{trilin-aaa-funct} with \eqref{trilin-ghost-funct}, it can be easily realized that the relevant part of $W(J,\eta,\bar\eta)$ corresponding to the gauge potential loop contribution to the 1-point function satisfies
\begin{equation}
W^{A}_1((J_3))=-2W^{\phi\pi}_1((J_3))\label{gaugevsghost}
\end{equation}
so that
\beqa
W_1((J_3))&:=&W^{A}_1((J_3))+W^{\phi\pi}_1((J_3))\nn\\
&=&\frac{k}{4 \lambda}\sum_{j,m,n,p,k,l}(j+1) (n-m)(P^{j,j}_{mp;nm}P^{j,j}_{pn;kl}-P^{j,j}_{pn;nm}P^{j,j}_{mp;kl})(\widetilde J_3)^j_{kl} \label{w1-gauge}.
\eeqa
The 1-point function with external $A_3$ is then given by
\be
\Gamma^{1}(A^3):=\sum_{j,n,p}\sigma^{j}_{3\ np}(A_3)^j_{np}=\frac{k}{ \lambda}\sum_{j,m,n,p}(j+1)(m-p)P^{j,j}_{mp;nm}(A_3)^j_{np}\label{effectgamma1}
\ee
\be
\sigma^{j}_{3\ np}=\frac{k}{\lambda}\sum_{-j\le m\le j}(j+1)(m-p)P^{j,j}_{mp;nm}=\frac{k}{\lambda}\delta_{np}\sum_{-j\le m\le j}(j+1)(m-n)P^{j,j}_{mp;nm}\label{1-pointa3}
\ee
where again $n,\ p$ (in addition to $j$) are external indices and the last relation in \eqref{1-pointa3} stems from the standard ''selection rule'' for the Clebsch-Gordan coefficients
\begin{equation}
m_1+m_2\ne k\Rightarrow \cg {j_1}{m_1}{j_2}{m_2}{l}{k}=0\label{selectrule},
\end{equation}
combined with \eqref{fuzzydag} and\eqref{matrixelemfuzzy}, implying that $n\ne p\Rightarrow P^{j,j}_{mp;nm}=0$. Despite the fact that  the propagator \eqref{propagator} has an IR singularity (at $l=0$) reflecting the fact that the classical gauge theory is massless by construction, the 1-point function $\sigma^{j}_{3\ np}$ \eqref{1-pointa3} has no IR singularity. Indeed, write $\sigma^{j}_{3\ np}$ as
\begin{eqnarray}
\sigma^{j}_{3\ np}&=&\frac{\lambda}{(2j+1)}\delta_{np}\big(\lim_{\varepsilon\to0}\frac{1}{\varepsilon}\sum_{-j\le m\le j}(m-n)
(Y^{j\dag}_{00})_{nm}(Y^j_{00})_{mn}\nonumber\\
&+&\sum_{m,l\ne0,k}\frac{(m-n)}{l(l+1)}(Y^{j\dag}_{lk})_{nm}(Y^j_{lk})_{mn}\big)\label{ir-regul}
\end{eqnarray}
where \eqref{propagator} has been used and we have introduced an IR regulator to parametrize the singularity. The first term in the parenthesis vanishes trivially since $(Y^j_{00})_{mn}\varpropto\delta_{mn}$ $(Y^j_{00}\varpropto\bbbone)$ holds true. Finally, summing all the contributions, $\sigma^{j}_{3\ np}$ can be written as
\begin{eqnarray}
\sigma^{j}_{3\ np}&=&(-1)^{2j}{\lambda}\delta_{np}\sum_{m,l\ne0,k}\frac{k}{l(l+1)}\cg {j}{m}{j}{-n}{l}{k}^2\nonumber\\
&=&(-1)^{2j}{\lambda}\delta_{np}\sum_{m,l\ne0}\frac{1}{l(l+1)}\sum_{k=-l}^lk\cg {j}{m}{j}{-n}{l}{m-n}^2\nonumber\\
&=&0          \label{1-pointa3fin},
\end{eqnarray}
where we used the ''selection rule'' \eqref{selectrule} to obtain the second relation in \eqref{1-pointa3fin}. \par 
Before computing the other components of the 1-point function, we notice that each of the ghost and gauge potential loop contributions to $\sigma^{j}_{3\ np}$ are separately vanishing. This comes from the combination of the derivative nature of the trilinear couplings (producing the factor $(m-n)$ in \eqref{sintghost}, \eqref{sintaaa}) together with the indices conservation law for the propagator \eqref{consindice2} which is reflected in the Clebsch-Gordan selection rule \eqref{selectrule}. We point out that the occurrence of the derivative in the trilinear couplings is essential to obtain the vanishing of $\sigma^{j}_{3\ np}$. In this respect, the appearance of a non zero tadpole at the 1-loop order can be expected in a scalar $\phi^3$ NCFT on $\mathbb{R}^3_\lambda$.\par 

The computation of the other components of the tadpole function can be conveniently carried out by setting $A_\pm=A_1\pm i A_2$ and  $D_\pm=\frac{i}{\lambda^2}[x_\pm, .]$.
Let us focus on the tadpole for $A_-$ (the analysis is similar for $A_+$). One has \cite{vit-wal-12}
\begin{eqnarray}
[x_+,v^j_{mn}]&=&\lambda\Bigl(F^j(m)v^j_{m+1,n}-F^j(n-1)v^j_{m,n-1}\Bigr),\ \forall j\in\frac{\mathbb{N}}{2},\ -j\le m,n\le j,\label{xminusaction}\\
F^j(m)&=&[(j+m+1)(j-m)]^{\frac{1}{2}}.
\end{eqnarray}
The relevant ghost-gauge interaction term is given by
\begin{eqnarray}
&S&_{int}^{A_-\phi\pi}=\frac{1}{2}\widetilde\tr(D_+\bar C[A_-,C])\nonumber\\
&=&\frac{k}{2\lambda}\sum_{j,m,n,p}(j+1) \Bigl (F^j(m)\bar C_{mn}(A_-)^j_{np}C_{pm+1}-F^j(n-1)\bar C_{mn}(A_-)^j_{n-1p}C_{pm}\Bigr.\nonumber\\
&-&\Bigl.F^j(m)\bar C_{mn}C_{np}(A_-)^j_{pm+1}+F^j(n-1)\bar C_{mn}C_{n-1p}(A_-)^j_{pm}\Bigr) \label{sintghostam}.
\end{eqnarray}
Hence
\begin{eqnarray}
S_{int}^{A_-\phi\pi}(\frac{\delta}{\delta (\widetilde J_+)},\frac{\delta}{\delta\widetilde{\bar\eta}},\frac{\delta}{\delta\widetilde\eta})&=&-\frac{k}{2\lambda}\sum_{j,m,n,p}(j+1)
\Bigl(F^j(m)\frac{\delta}{\delta\widetilde{\bar\eta}_{m+1p}}\frac{\delta}{\delta (\widetilde{J}_+)_{pn}}\frac{\delta}{\delta\widetilde{\eta}_{nm}}\nonumber\\
&-&F^j(n-1)\frac{\delta}{\delta\widetilde{\bar\eta}_{mp}}\frac{\delta}{\delta (\widetilde J_+)_{pn-1}}\frac{\delta}{\delta\widetilde\eta_{nm}}
-F^j(m)\frac{\delta}{\delta\widetilde{\bar\eta}_{pn}}\frac{\delta}{\delta\widetilde\eta_{nm}}\frac{\delta}{\delta (\widetilde J_+)_{m+1p}}\nonumber\\
&+&F^j(n-1)\frac{\delta}{\delta\widetilde{\bar\eta}_{pn-1}}\frac{\delta}{\delta\widetilde\eta_{nm}}\frac{\delta}{\delta (\widetilde J_+)_{mp}}\Bigr)\label{trilin-app-funct}
\end{eqnarray}
where we have defined
\begin{equation}
(\widetilde J_\pm):=\frac{1}{2}(\widetilde J_1\pm i\widetilde J_2).
\end{equation}
Thus, one has
\begin{equation}
W_0(J,\eta,\bar\eta):=\sum\Bigl( \frac{1}{2}(\widetilde J_+)^{j_1}_{mn}P^{j_1 j_2}_{mn;kl} (\widetilde J_-)^{j_2}_{kl}+\frac{1}{8}(\widetilde J_3)^{j_1}_{mn}P^{j_1 j_2}_{mn;kl} (\widetilde J_3)^{j_2}_{kl}+\widetilde{\bar\eta}^{j_1}_{mn}P^{j_1 j_2}_{mn;kl}\widetilde \eta^{j_2}_{kl}\Bigr)\label{formalW0prime}
\end{equation}
From \eqref{trilin-app-funct} and \eqref{1loopgeneric}, the relevant part of $W(J,\eta,\bar\eta)$ related to the 1-loop ghost contribution to the tadpole function for $A_-$ is
\begin{eqnarray}
W^{\phi\pi}_1((J_\pm))&=&\frac{k}{4\lambda}\sum_{j,m,n,p}\Bigl(F^j(m)P^{jj}_{m+1p;nm}P^{jj}_{pn;kl}-F^j(n-1)P^{jj}_{mp;nm}P^{jj}_{pn-1;kl}\nonumber\\ &-&F^j(m)P^{jj}_{pn;nm}P^{jj}_{m+1p;kl}+F^j(n-1)P^{jj}_{pn-1;nm}P^{jj}_{mp;kl}\Bigr)(\widetilde J_-)_{kl}.
\end{eqnarray}
From this follows
\begin{eqnarray}
\Gamma^{1\ \phi\pi}(A_-):&=&\sum_{j,n,p}\sigma^{j\ \phi\pi}_{+\ np}(A_-)^j_{np}\nonumber\\
&=&\frac{k}{2\lambda}\sum_{j,m,n,p}(j+1)\Bigl(F^j(m)P^{jj}_{m+1p;nm}(A_-)^j_{np}-F^j(n-1)P^{jj}_{mp;nm}(A_-)^j_{n-1p}\nonumber\\ &-&F^j(m)P^{jj}_{pn;nm}(A_-)^j_{pm+1}+F^j(n-1)P^{jj}_{pn-1;nm}(A_-)^j_{pm}\Bigr),
\end{eqnarray}
and finally
\begin{eqnarray}
\sigma^{j\ \phi\pi}_{+\ np}&=&\frac{k}{2\lambda}\sum_{m=-j}^j(j+1)\Bigl(F^j(m)P^{jj}_{m+1p;nm}-F^j(n)P^{jj}_{mp;n+1m}\nonumber\\
&-&F^j(p-1)P^{jj}_{nm;mp-1}+F^j(m-1)P^{jj}_{nm-1;mp}\Bigr)\label{1point-aminus}.
\end{eqnarray}
Further using \eqref{selectrule}which encodes \eqref{consindice2}, Eq. \eqref{1point-aminus} can be expressed as
\begin{eqnarray}
\sigma^{j\ \phi\pi}_{+\ np}&=&\frac{k}{2\lambda}\delta_{n+1,p}\Pi(j,n),\label{tadpole-aminus}\\
\Pi(j,n)&=&\sum_{m=-j}^j(j+1)\big(2F^j(m)P^{jj}_{m+1n+1;nm}-F^j(n)P^{jj}_{mn+1;n+1m}-F^j(n)P^{jj}_{mn;nm}\big)\label{goldstoneoperator}.
\end{eqnarray}
Next, by inspection of the interaction term for $A_i$, it can be realized that the gauge potential loop contribution satisfies a relation similar to \eqref{gaugevsghost} so that the tadpole function for $A_+$ defined by
\begin{equation}
\Gamma^{1}(A_+):=\sum_{j,n,p}\sigma^{j}_{-\ np}(A_+)^j_{np}
\end{equation}
is given by
\begin{equation}
\sigma^{j\ \phi\pi}_{-\ np}=-\frac{k}{2\lambda}\delta_{n,p+1}\Pi(j,n)\label{tadpole-aplus}, 
\end{equation}
where $\Pi(j,n)$ is stil given by \eqref{goldstoneoperator}.\par

It is convenient to introduce the Wigner $3j$ symbols
\begin{equation}
\cg{j_1}{m_1}{j_2}{m_2}{j_3}{m_3}=(-1)^{j_2-j_1-m_3}{\sqrt{2j_3+1}}\wign{j_1}{m_1}{j_2}{m_2}{j_3}{-m_3}\label{cg-wign}
\end{equation}
and express the propagator \eqref{propagator} as
\begin{equation}
P^{jj}_{mn;pq}=\frac{\lambda^2}{k(j+1)}\sum_{l,k}(-1)^{-(n+p)}\frac{2l+1}{l(l+1)}\wign{j}{m}{j}{-n}{l}{k}\wign{j}{q}{j}{-p}{l}{k}\label{practice-propa}.
\end{equation}
Then, by using the orthogonality relations on the Wigner $3j$ symbols, one can show that
\begin{equation}
\sum_{m}P^{jj}_{mn;nm}=\frac{\lambda^2}{k(j+1)}\frac{(-1)^{2n}}{2j+1}\sum_{l=0}^{2j}\frac{2l+1}{l(l+1)}\label{sum-boundary},
\end{equation}
which permits to sum up the last two terms in \eqref{goldstoneoperator}, leading to
\begin{equation}
\Pi(j,n)=2\sum_{m=-j}^j(j+1)\big(F^j(m)P^{jj}_{m+1n+1;nm}-F^j(n)P^{jj}_{mn;nm}\big)\label{goldstoneoperator1}.
\end{equation}

The tadpoles \eqref{tadpole-aminus}, \eqref{tadpole-aplus} have no IR singularity, as it happens for $\sigma^{j}_{3\ np}$ \eqref{1-pointa3}. Indeed, by isolating the IR singularity of the propagator as in \eqref{ir-regul}, using in particular $(Y^j_{00})_{mn}\varpropto\delta_{mn}$, one extracts readily from \eqref{goldstoneoperator1} the potentially IR singular part of the tadpoles given by
\begin{equation}
\Pi(j,n)_{IR}\sim\lim_{\varepsilon\to0}\frac{1}{\varepsilon}\sum_m(F^j(m)\delta_{nm}-F^j(n)\delta_{nm})=0
\end{equation}
so that $\Pi(j,n)$ is IR finite.\par 

Let us summarize our computation. In the diagonal gauge defined by \eqref{special-gauge}, we find that the one-loop effective action related to the BRST invariant gauge-fixed theory \eqref{gaugefixfinal} is given by
\begin{equation}
\Gamma^1(A_i)=\frac{k}{2\lambda}\sum_{j\in\frac{\mathbb{N}}{2}}\sum_{-j\le n,p\le j}\Pi(j,n)(\delta_{n+1,p}(A_-)^j_{np}-\delta_{n,p+1}(A_+)^j_{np})\label{effective-lin}
\end{equation}
with $\Pi(j,n)$ given by \eqref{goldstoneoperator}.

\section{Discussion}\label{discussion}

It turns out that the $\sigma^j_{\pm\ np}$'s are not all identically zero, unlike $\sigma^j_{3\ np}$ \eqref{1-pointa3fin}. Indeed, by computing for instance $\Pi(j,n)$ for $j=\frac{1}{2}$, $n=-\frac{1}{2}$, one finds that $\Pi(\frac{1}{2},-\frac{1}{2})=2$ which signals the non-vanishing of the 1-point function, i.e terms linear in $A_i$, albeit absent in the classical action by the very construction, are generated again by quantum fluctuations in the effective action.\par

It is instructive to characterize the UV behavior of the tadpole function. This task is complicated by the structure of \eqref{goldstoneoperator1}  
 by the various summations that are entangled. One has to analyze separately the cases $j\to\infty$ with finite $n$, say $n\ll j$, and $j\to\infty$ with $\vert n\vert=j$. This can be achieved by using the standard 3-terms recursion relation among Clebsch-Gordan coefficients (and consequently fuzzy spherical harmonics). The technical details are given in the appendix \ref{appendix2}. From the formula \eqref{limit-pi}, one infers that the large $j$ limit of the tadpole can be infinite, namely
\begin{equation}
\lim_{j\to\infty}\Pi(j,-j)\sim-{\lambda}{\sqrt{j}}\log j.\label{limit-effectv}
\end{equation}
By further using \eqref{x0v}
\begin{equation}
x_+v^j_{mn}=\lambda F^j(m)v^j_{m+1,n},\ x_-v^j_{mn}=\lambda F^j(m-1)v^j_{m-1,n}\label{shifter}
\end{equation}
it can be realized that \eqref{effective-lin} can be recast into the form
\begin{equation}
\Gamma^1(A_i)=\frac{1}{2}\widetilde\tr(\eta_+{\hat{A}}_-+\eta_-{\hat{A}}_+)\label{renor-wavefunct1},
\end{equation}
(recall $\eta_i=\frac{i}{\lambda^2}x_i$) where the field ${hat{A}}_\pm$ are defined from their expansion coefficients with
\begin{equation}
({\hat{A}}_-)^j_{n,n+1}=\frac{\Pi(j,n)}{(j+1)F^j(n)}(A_-)^j_{n,n+1},\ ({\hat{A}}_+)^j_{n,n-1}=\frac{\Pi(j,n)}{(j+1)F^j(n-1)}(A_+)^j_{n,n-1}\label{z-wave}
\end{equation}
while the other components $({\hat{A}}_\pm)^j_{n,p},\ p\ne n\mp 1$ do not appear in \eqref{renor-wavefunct1} which is simply due to \eqref{shifter} and \eqref{matrixprod}, \eqref{properties2b}. Notice that the factors $\frac{\Pi(j,n)}{F^j(n)}$ and $\frac{\Pi(j,n)}{F^j(n-1)}$ affecting the expansion modes in \eqref{z-wave} have a vanishing large $j$ limit thanks to \eqref{limit-effectv}. The occurrence of a one-loop non vanishing 1-point function and its particular expression within the massless gauge theory we have considered, eqns. \eqref{renor-wavefunct1} and \eqref{z-wave}, has interesting consequences that we discuss now. \par 

Eqn. \eqref{z-wave} can be interpreted as a wave function renormalisation of each of the (lower and upper first subdiagonal) modes of $A_\pm$ in its expansion in the canonical basis. It is somewhat different from what would usually happen in commutative field theories for which wave function renormalisation results in overall factors rescaling the fields.\par 

The occurrence of non-vanishing tadpole $\Gamma^1(A_i)$ \eqref{renor-wavefunct1} signals that the classical vacuum configuration becomes unstable under quantum fluctuations. Thus, moving ahead consistently into the perturbative expansion would need to tune the vacuum at each order, i.e performing an expansion of the field variable around the right vacuum at each order of perturbation, leading presumably to a massive theory.\par 

Next, eqn. \eqref{renor-wavefunct1} shows that some but not all the terms linear in $A_i$, $\sim\widetilde\tr(\eta_i A_i)$, that were absent from the classical action by construction are restored at one-loop order. Only $A_3$ has vanishing tadpole function. Notice by the way that this result, obtained in the diagonal gauge \eqref{special-gauge}, obviously holds true in the axial gauge $A_3=0$. Besides, one observes that $\Gamma^1(A_i)$ does not have the standard expression of the $\sigma$-term of the old linear $\sigma$-models. Namely, $\Gamma^1(A_i)\ne\widetilde\tr(\sigma_i A_i)\sim\sum\sigma_i (A_i)^j_{mm}$ which involves only the ''diagonal modes''. Instead, it has the ''non-covariant'' form \eqref{renor-wavefunct1} which may be interpreted as an explicit breaking term of the global rotational invariance of the effective action. \par 

Keeping in mind \eqref{actionansatz}, \eqref{condition2bis}, the above discussion suggests the appearance at one-loop of a mass splitting for the triplet $({\cal{A}}_i),\ i=1,2,3$ with corresponding term in the effective action given by $\sum_{k=1}^3\mu_k{\cal{A}}_k^2$, replacing the term $\sim\widetilde\tr(\mu{\cal{A}}_i{\cal{A}}_i)$ in \eqref{actionansatz}. At the classical order, the relation \eqref{condition2bis} insures the vanishing of the tadpole (linear terms) for each of the $A_i$ with $\mu_3=\frac{3}{2\lambda}\gamma=\mu_1=\mu_2$ and the triplet has a ''mass degeneracy''. This latter is removed by quantum fluctuations and thus the relation $\mu_3=\frac{3}{2\lambda}\gamma$ would still hold true among renormalized parameters insuring that $\Gamma^1(A_3)=0$ while $\mu_1,\mu_2\ne \mu_3$ with $\Gamma^1(A_i)\ne0$, $i=1,2$.\par 

To summarize, we have considered a wide class of gauge invariant models on the noncommutative space $\mathbb{R}^3_\lambda$, stemming from a natural differential calculus based on derivations of $\mathbb{R}^3_\lambda$ and assuming a noncommutative analog of the Koszul notion of connection. The related curvature, upon squaring, gives rise generally to mass terms for the gauge potential $A_i$. In order to mimic salient classical features of commutative Yang-Mills theory, we focused on models which are massless and with no linear $A_i$ dependence. This yields to noncommutative gauge models for which the propagator can be computed in a convenient gauge that may be viewed as an analog of the covariant gauges used within commutative gauge theories. Working in this gauge, we have found that the infrared singularity of the propagator stemming from masslessness disappears from the computation of the correlation functions. We have shown that massless gauge invariant models on $\mathbb{R}^3_\lambda$ have quantum instabilities of the vacuum, signaled by the occurrence of non vanishing tadpole (1-point) functions for some but not all of the components of the gauge potential. It appears that the tadpole contribution to the effective action cannot be interpreted as a standard $\sigma$-term while its global symmetry does not fit with the one of the classical action, akin to an explicit symmetry breaking term. It would be interesting to examine whether this can be actually related to some kind of analog of radiative pseudo-Goldstone mechanism. This will be examined in a forthcoming publication.\par 

\vskip 2 true cm
{\bf{Acknowledgments}}: We thank H. Steinacker for useful correspondence on the relationship between a special truncation the gauge model considered here and the brane model introduced in \cite{ARS00}. A. G. is grateful to N. Drago, T.-P. Hack and N. Pinamonti for useful discussions. J.-C. W thanks D.N. Blaschke, M. Dubois-Violette and H. Grosse for discussions at various stages of this work and the hospitality of the Dipartimento di Fisica, Universit\`a di Napoli Federico II and INFN, Sezione di Napoli where a part of this work has been done.

\setcounter{section}{0}
\appendix

\section{\texorpdfstring{General properties of $\mathbb{R}^3_\lambda$ and related matrix bases}{Diff calcul}}\label{appendix0}
In this appendix we briefly review the derivation of the algebra $\Rl$ and its matrix basis \cite{vit-wal-12}.
 It can be viewed as a particular quadratic subalgebra of $(\mathbb{R}^4_\theta,\ \star_V)$, the associative algebra of functions of $\mathbb{R}^4\simeq\mathbb{C}^2$.   $\star_V$ is the  Wick-Voros product \cite{Wick-Voros}, a variation of the Moyal product, given by
\begin{equation}
\phi\star_V \psi\, (z_a,\bar z_a)= \phi(z,\bar z) \exp(\theta \overleftarrow\del_{z_a}\overrightarrow\del_{\bar z_a}) \psi(z,\bar z), \,\,\,\, a=1,2 \label{Wick-Vorosprod}
\end{equation}
with  $\theta$ a constant parameter, which, differently from \cite{vit-wal-12}, we choose here to have length dimension 1, while $z^a, \bar z^a$  coordinate functions on $\mathbb{C}^2$ have length dimension 1/2. 
Denoting by $(x^{\mu=0,1,2,3})$ the subalgebra of quadratic functions  of $\mathbb{R}^4_\theta$ defined by
\begin{equation}
x^\mu=  \bar z^a \frac{\sigma^\mu_{ab}}{2} z^b, \;\;\; \mu=0, ..,3 \label{xmu}\end{equation}
with $\sigma^i$
  the Pauli matrices, and $\sigma^0= \mathbf{1}_2$, 
the set of polynomial functions of $x^\mu$  is a subalgebra  with respect to the Wick-Voros product, so that a new product gets  induced in the subalgebra
\begin{equation}
\phi\star \psi \,(x)= \exp\left[\frac{\lambda}{2}\left(\delta^{ij} x^0+ i \epsilon^{ij}_k x^k \right)\frac{\del}{\del u^i}\frac{\del}{\del v^j}\right] \phi(u) \psi(v)|_{u=v=x} \label{astarsu2}
\end{equation}
where $\lambda=\theta$ is the  noncommutative parameter of length dimension 1. Let us notice that $(x^0)^2= \sum_i (x^i)^2$  so that in the commutative limit,  this subalgebra, which we identify as $\Rl$,  yields back the algebra of functions on $\mathbb{R}^3$. Moreover, $x^0$ $\star$-commutes with all elements of the subalgebra, so that we can alternatively define 
 $\Rl$ as the $\star$-commutant of $x_0$ in $\mathbb{R}^4_\theta$. The star product \eqref{astarsu2} implies for coordinate functions
 \begin{eqnarray}
x^i\star x^j&=& x^i x^j+ \frac{\lambda}{2} \left(x^0 \delta^{ij} + i \epsilon^{ij}_k x^k\right);\ x^0\star x^i = x^i\star x^0 = x^0 x^i + \frac{\lambda}{2} x^i;\\
x^0\star x^0&=&(x^0)^{*2}=x^0(x^0+\frac{\lambda}{2}) = \sum_{i=1}^3 x^i\star x^i- \lambda x^0 . \label{ax0*2}
 \end{eqnarray}
from which one obtains
\begin{equation}
 [x^i,x^j]_\star=i \lambda \epsilon^{ij}_k  x^k \label{acommsu2}.
\end{equation}
Thus,  $\mathbb{R}^3_\lambda$ can be viewed as $\mathbb{R}^3_\lambda=\mathbb{C}[x^0,x^i]/{\cal{I}}_{{\cal{R}}_1,{\cal{R}}_2 }$, the quotient of the free algebra generated by the coordinates $(x^\mu)_{\mu=0,1,2,3}$ by the two-sided ideal generated by the relation ${\cal{R}}_1: [x^i,x^j]_\star=i\lambda\epsilon^{ij}_k x^k$, together with ${\cal{R}}_2: x^0\star x^0 +\lambda x^0 = \sum_i x^i\star x^i$. Finally, notice that ${\cal{U}}(\mathfrak{su}(2))\subset\mathbb{R}^3_{\lambda\ne 0}$, where ${\cal{U}}(\mathfrak{su}(2))$ denotes as usual the universal enveloping algebra of the Lie algebra $\mathfrak{su}(2)$.\par 

Let us now shortly review  the matrix basis adapted to $\Rl$ constructed in \cite{vit-wal-12, duflo}. It is obtained by reduction of the Wick- Voros matrix basis for $\mathbb{R}^4_\theta$ \cite{discofuzzy}. The two-dimensional Wick Voros basis (of which the four dimensional one is a straightforward  extension)  is defined in terms of the  weighted quantization map
\begin{equation}
\hat \phi:=\hat {\mathcal{W}}_V(\phi)=\frac{1}{(2\pi)^2}\int\dd^2 z\  \dd^2 \eta\, e^{{ -( \eta \bar z-\bar\eta z )}} e^{\theta\eta a^\dag} e^{-\theta\bar\eta a }
\phi(z, \bar z);\    [a, a^\dag]=\theta\label{aWeylmap}
\end{equation}
for any well-behaving function $\phi(z,\bar{z})$ on $\mathbb{C}$, where $a, a^\dag$ are creation
and annihilation operators acting on ${\cal{H}}_0\cong \ell^2(\mathbb{N})$, the Hilbert space of the 1-d harmonic oscillator with orthonormal basis $(|n\rangle)_{n\in\mathbb{N}}$. The inverse map and the defining relation for $\star_V$  (of which an asymptotic form is represented by Eq. \eqref{Wick-Vorosprod}) are
\begin{equation}
\phi(z,\bar z) = \mathcal{W}_V^{-1}(\hat \phi)= \langle z|\hat \phi|z\rangle,\ \phi\star_V \psi := \mathcal{W}_V^{-1}\left(\hat{\mathcal{W}}_V(\phi)\hat {\mathcal{W}}_V(\psi)\right)= \langle z|\hat \phi\,\hat \psi|z\rangle
\label{Wick-Voros}
\end{equation}
where $|z\rangle$ are {\it coherent states} defined by $a|z\rangle=z|z\rangle$. For more details, see e.g \cite{discofuzzy}. Then, \eqref{aWeylmap} associates to  analytic functions  normal ordered operators of the form $\hat \phi =\hat {\mathcal{W}}_V(\phi)=\sum_ {pq}\tilde\phi_{pq}  a^{\dag p} a^q $. 

The extension to the 4-dimensional case is easily achieved by considering two pairs of operators $(a_a,\ a^\dag_a)$, $a=1,2$, each one acting on one copy of ${\cal{H}}_0$.  We thus have the Hilbert space  ${\cal{H}}={\cal{H}}_0\otimes {\cal{H}}_0$ of  two 1-dimensional harmonic oscillators with orthonormal basis $|N\rangle:= |n_1\rangle \otimes |n_2\rangle , n_1, n_2\in\mathbb{N}$.  Then, the Bargman-Jordan-Schwinger realization of $\mathfrak{su}(2)$ leads to a natural basis in  $\mathbb{R}^4_\theta$:
\begin{equation}
\{\hat v^{j\tilde\jmath}_{m\tilde m}:=|j,m\rangle\langle \tilde\jmath,\tilde m|\},\ j,\tilde\jmath\in{{\mathbb{N}}\over{2}},\ -j\le m\le j\, , -\tilde\jmath\le \tilde m\le \tilde\jmath \label{2sphericbasis}\, .
\end{equation}
where $\ |j,m\rangle$ is a short-hand for $|j+m\rangle\otimes| j-m\rangle$  and the relation to the oscillators basis is simply furnished by $j+m=n_1, j-m=n_2$.  In other words, we are using the well known canonical  decomposition of ${\cal{H}}$ as   ${\cal{H}} = \bigoplus_{ j\in{{\mathbb{N} }\over{2}}  }{\cal{V}}_j$ where
\begin{equation}
{\cal{V}}_j={\rm span}\;\{ |j,m\rangle\}_{-j\le m\le j},\ |j,m\rangle:=|j+m\rangle\otimes| j-m\rangle\label{irrepssu2}
\end{equation}
is the linear space carrying the irreducible representation of $SU(2)$ with dimension $2j+1$. For any $j\in{{\mathbb{N} }\over{2}} $, the 
system $\{ |j,m\rangle\}_{-j\le m\le j}$ is orthonormal. Any function of $\mathbb{R}^4_\theta$ can then be expanded in terms of the symbols of the operators $\hat v^{j\tilde\jmath}_{m\tilde m}$ of Eq.  \eqref{2sphericbasis} given by $ v^{j\tilde\jmath}_{m\tilde m}(z_a,\bar z_a)=\langle z_1,z_2\vert\hat v^{j\tilde\jmath}_{m\tilde m}\vert z_1,z_2 \rangle$. 
On imposing that they $\star$-commute with $x^0$
\begin{equation}
x^0 \star_V v^{j\tilde\jmath}_{m\tilde m} (z,\bar z)-v^{j\tilde\jmath}_{m\tilde m}\star_V
x^0 (z,\bar z)=\lambda(j-\tilde\jmath) v^{j\tilde\jmath}_{m\tilde m}
\end{equation}
we finally deduce that a basis for $\Rl$ (regarded as an algebra of operators)  is represented   by the family of operators
\begin{equation}
\{\hat v^{j}_{m\tilde m}:=\hat v^{jj}_{m\tilde m}=|j,m\rangle\langle j,\tilde m|\},\ j\in{{\mathbb{N}}\over{2}},\ -j\le m\le j\, , -\tilde j\le \tilde m\le \tilde j \label{basisr3l1}\,.
\end{equation}
Equivalently, when regarding     $\Rl$ as a noncommutative algebra of functions with the  star product \eqref{starsu2},   the matrix basis will be  given by the symbols 
\begin{equation}
v^{j}_{m\tilde m}(z_a,\bar z_a)=\langle z_1,z_2\vert\hat v^{j}_{m\tilde m}\vert z_1,z_2 \rangle = e^{-\frac{\bar z_a z_a}{\theta}}\frac{\bar z_1^{j+m}
z_1^{j+\tilde m} \bar z_2^{j-m}
z_2^{j-\tilde m}}{\sqrt{(j+m)!(j-m)! (j+\tilde m)!(j-\tilde m)! \theta^{4j} }} \label{matrixbasis}
\end{equation}
which can be expressed in terms of the coordinates $x^\mu$ (although not uniquely) as in Eq. \eqn{xmatrixbasis}.
For any function in $\Rl$ we have then
 \begin{equation}
\phi(x)=\sum_{j}\sum_{m,\tilde m=-j}^j \phi^j_{m\tilde m} v^j_{m\tilde m}(x).
\label{phixi1}
\end{equation}
The following properties hold true:
\begin{equation}
\hat v^{j_1}_{m_1,m_2}\hat v^{j_2}_{n_1,n_2} =\delta^{j_1j_2}\delta_{m_2n_1}\hat v^{j_1}_{m_1,n_2},\ {{(\hat v^{j}_{m_1,m_2})^\dag}}=\hat v^{j}_{m_2,m_1}
\label{properties2a}
\end{equation}
with equivalent expressions for their symbols (cfr. Eq \eqn{matrixprod}). 
The $\star$-product on $\mathbb{R}_\lambda^3$, Eq. \eqref{starsu2}  reduces  then to a blockwise diagonal matrix product
\begin{eqnarray}
\phi\star \psi (x)&=&\sum \phi^{j}_{m_1\tilde m_1} \psi^{j}_{m_2\tilde m_2}
v^{j}_{m_1\tilde m_1} \star v^{j}_{m_2\tilde m_2} \; =\sum \phi^{j}_{m_1\tilde m_1}
\psi^{j}_{m_2\tilde m_2} v^{j}_{m_1\tilde m_2} \delta_{\tilde m_1 m_2}\nonumber\\
&=& \sum_{j,m_1, \tilde m_2} (\Phi^j\cdot \Psi^j)_{m_1 \tilde m_2} v^j_{m_1 \tilde m_2}
\label{starmatrix1}
\end{eqnarray}
where we have introduced the  infinite, block-diagonal matrix $\Phi$, each block being the $(2j+1)\times (2j+1)$ matrix   $\Phi^j=\{\phi^j_{mn}\}, \, -j\le m,n\le j$.
Thus we compute \cite{vit-wal-12}
\begin{equation} \label{x0v}
\begin{array}{lll}
x_+\star v^j_{m\tilde m}= \lambda \sqrt{(j+m+1)(j-m)} v^j_{m+1 \, \tilde m}  & & v^j_{m\tilde m}\star x_+= \lambda\sqrt{(j-\tilde m+1)(j+\tilde m)} v^j_{m\, \tilde m -1} \\
x_-\star v^j_{m\tilde m}= \lambda \sqrt{(j-m+1)(j+m)} v^j_{m-1\,\tilde m}  &&v^j_{m\tilde m}\star x_-= \lambda\sqrt{(j+\tilde m+1)(j-\tilde m)} v^j_{m \,\tilde m +1} \\
x_3\star v^j_{m\tilde m}= \lambda\, m \, v^j_{m\tilde m}&&  v^j_{m\tilde m} \star x_3= \lambda \,\tilde m \, v^j_{m\tilde m}\\
x_0\star v^j_{m\tilde m}= \lambda\, j \, v^j_{m\tilde m}&&  v^j_{m\tilde m} \star x_0= \lambda\, j \,v^j_{m\tilde m}
\end{array}
 \end{equation}
were we have introduced
\begin{equation}
x_\pm:= x_1\pm i x_2.
\end{equation}

%To summarize, we have that, for any ${j\in { {{ {\mathbb{N}} 
%}\over{2} }}}$, the $(2j+1)^2$ the linear maps $\hat v^j_{m_1,m_2}:{\cal{V}}^j\to{\cal{V}}^j$, $-j\le m_1,m_2\le j$  determine %the canonical basis of the algebra of endomorphisms of ${\cal{V}}^j$, $End({\cal{V}}^j)$, $\forall {j\in { {{ {\mathbb{N}} } 
%\over{2} }}}$, i.e the so-called fuzzy sphere ${\mathbb{S}}^j$ of radius related to $j$. The canonical basis is obviously %orthonormal with respect to the scalar product defined by \eqref{scalarproduct}. Besides, $\mathbb{R}^3_\lambda$ is %diffeomorphic to  the direct sum decomposition
%\begin{equation}
%\mathbb{R}^3_\lambda\simeq\bigoplus_{j\in { {{ {\mathbb{N}} }\over{2} }}}End({\cal{V}}^j) \simeq\bigoplus_{j\in 
%{ {{ {\mathbb{N}} }\over{2} }}}\mathbb{S}^j\label{sumfuzzy1}.
%\end{equation}
%For more details, see \cite{vit-wal-12}. 

Expressing the fields of NCFT on $\Rl$ in the canonical basis yields diagonal interaction vertices so that this latter may be physically viewed as the interaction basis.

It is well known and widely used  in the  context of fuzzy spheres  that $End({\cal{V}}^j)$
 is spanned by  the so called Fuzzy Spherical Harmonics Operators, or, up to normalization factors,  irreducible tensor operators.
 We shall indicate them as
 \begin{equation}
 \hat Y^j_{lk}\in End({\cal{V}}^j),\;\; l\in\mathbb{N},\;\; \ 0\le l\le 2j,\;\; -l\le k\le l,
  \end{equation}
  whereas the unhatted objects $Y^j_{lk}$ are their symbols and are sometimes referred to as fuzzy spherical harmonics with no other specification (notice however
that the  functional form  of the symbols does depend on the dequantization map that has been chosen).
Concerning  the definition and normalization of the fuzzy spherical harmonics operators, we use the following conventions \cite{Das}. We set
 \begin{equation}
  J_{\pm}:=\frac{{\hat x}_{\pm}}{\lambda}; ~~~~~~J_3:=\frac{{\hat x}_{3}}{\lambda} .
 \end{equation}
We have, for $l=m$,
 \begin{equation}
\hat Y^j_{ll}:=(-1)^l  \frac{ \sqrt{2j+1}}{ l !}
\frac {\sqrt{ (2l+1)! (2j-l)! } } { (2j+l+1)!}  (J_+)^l   \label{highestfuzzyharmonics}
\end{equation}
while the others  are defined recursively through the action of $J_-$
 \begin{equation}
\hat Y^j_{lk}:= [ (l+k+1) (l-k)]^{-\frac{1}{2}} [J_-,\hat Y^j_{l,k+1}],\;\;  \label{fuzzyharmonics}
\end{equation}
and satisfy
 \begin{equation}
(\hat Y^j_{lk})^\dag=(-1)^{k-2j} \hat Y^j_{l,-k},\
\langle \hat Y^j_{l_1 k_1},\hat Y^j_{l_2 k_2}\rangle=\tr((\hat Y^j_{l_1k_1})^\dag{ \hat Y}^j_{l_2k_2})=(2j+1)\delta_{l_1l_2}\delta_{k_1k_2} \label{relatfuzzyharm}.
\end{equation}
The symbols are defined through the dequantization map 
 \begin{equation}
Y^j_{lk}:=\langle z_1,z_2|\,\hat Y^j_{lk}\,|z_1,z_2\rangle. \label{fuzzyha}
\end{equation}
From\eqref{fuzzyharmonics}, \eqref{relatfuzzyharm} and the Lie algebra relation $[J_+,J_-]= 2J_3$ it is straightforward to check the usual properties
\begin{eqnarray}
{[} J_{-},{\hat Y}^j_{lk} { ] }&=& \sqrt{ (l+k)(l-k+1) } { \hat Y}^{j}_{l \, k-1}  \\
{[} J_{+},{\hat Y}^j_{lk} { ] }&=& \sqrt{(l-k)(l+k+1)} { \hat Y}^{j}_{l \, k+1} \\
{[}J_3,{\hat Y}^j_{lk} { ] } &=& k\; {\hat Y}^{j}_{l  k}\\
{[} J_{i},{[} J_{i},{\hat Y}^j_{lk} {]} {]}&=& l(l+1) {\hat Y}^{j}_{l  k}
\end{eqnarray}
which imply for the symbols
\begin{eqnarray}
{[}x_-,Y^j_{lk}{ ] }_\star=\lambda  <z|{[} J_{-},{\hat Y}^j_{lk} { ] }|z>= \lambda \sqrt{(l+k)(l-k+1)}Y^{j}_{l \, k-1} \nonumber\\
{[}x_+,Y^j_{lk}{ ] }_\star= \lambda  <z| {[} J_{+},{\hat Y}^j_{lk} { ] } |z>= \lambda \sqrt{(l-k)(l+k+1)}  Y^{j}_{l \, k+1}\nonumber\\
{[}x_3,Y^j_{lk}{ ] }_\star=\lambda  <z|{[}J_3,{\hat Y}^j_{lk} { ] } |z>=\lambda \; k \; Y^{j}_{l  k}\label{xharm}
\end{eqnarray}
and in particular
 \begin{equation}
{[}x_i,{[}x_i, Y^j_{lk}{ ]}_\star {]}_\star=\lambda ^2 <z|{[} J_{i},{[} J_{i},{\hat Y}^j_{lk} {]} {]} |z>=\lambda^2 \; l(l+1) \; Y^{j}_{l  k}. \label{xxharm}
\end{equation}

It turns out that a class of natural Laplacian operators on $\mathbb{R}^3_\lambda$, such as the one considered in \cite{vit-wal-12} is diagonal in this basis, together with the suitably gauge-fixed kinetic operator of the gauge models built in this paper. This basis may then be viewed physically as the propagation basis. \par 

The relation between the two  bases reads as follows 
\begin{equation}
 \hat Y^j_{lk}=\sum_{-j\le m,\tilde m\le j} ( Y^j_{lk})_{m \tilde m} \hat v^j_{m \tilde m},\   \;\;\; Y^j_{lk}=\sum_{-j\le m,\tilde m\le j} ( Y^j_{lk})_{m \tilde m} v^j_{m \tilde m},\ \label{matrixelemfuzzydef1}
\end{equation}
where the coefficients in the expansion can be written as
\begin{equation}
(Y^j_{lk})_{m \tilde m}=\langle \hat v^j_{m \tilde m}|\hat Y^j_{lk}\rangle={\sqrt{2j+1}}(-1)^{j-\tilde m}\cg {j}{m}{j}{-\tilde m}{l}{k},\ -j\le m,\tilde m\le j \, ,\label{matrixelemfuzzy}
\end{equation}
\begin{equation}
({Y^j_{lk}}^\dag)_{m \tilde m}=(-1)^{-2j} (Y^j_{lk})_{\tilde m m}. \label{fuzzydag}
\end{equation}

%In the subsequent analysis, it will be convenient to use the equivalent description of $\Rl$ as $\mathcal{W}_V(\Rl)$, i.e already represented on the operators of ${\cal{H}}$ via the map $\mathcal{W}_V$. Any element of $%\mathcal{W}_V(\Rl)$ is then given by the rightmost equality of \eqref{phixi} and the $\star$-product reduces to the matrix product \eqref{starmatrix}. 
%To simplify the notations, we set from now on $\mathcal{W}_V(\Rl)=\Rl$, %and drop the hat and $\star$ symbols. 
%To simplify the notations, we drop from now on the $\star$ product, whenever no ambiguity is possible. 

\section{Perturbative set-up}\label{appendix1}
The free part of the gauge sector of the theory is controlled by
\begin{equation}
Z_{f;A}(J)=\int{\cal{D}}{A}\ e^{-\big(S_2(A)+\widetilde\tr (A^j_i) (J^j_i) \big)}=\exp\big(\frac{1}{8}\sum(\widetilde J_i)^{j_1}_{mn}P^{j_1 j_2}_{mn;kl} (\widetilde J_i)^{j_2}_{kl}\big)\label{functionalfreegauge},
\end{equation}
where the the source variable $(J_i)^j_{nm}$ corresponds to the field variable $(A_i)^j_{mn}$. Moreover, we have redefined $\widetilde J^j = k (j+1) J^j$, with $k=8\pi\lambda^3/g^2$,  to take into account the weight-factor of the trace. The second relation{\footnote{The unessential prefactor in the 2nd relation is not explicitly written.}} is obtained from  $A^j_{mn}=A^{\prime j}_{mn}-\frac{1}{4}P^{jj}_{nm;kl}\widetilde J^j_{kl}$. 

For the (Grassmann) free ghost sector, one obtains
\begin{eqnarray}
Z_{f;\phi\pi}(\bar \eta,\eta)&=&\int{\cal{D}}{\bar C}{\cal{D}}C\ e^{-\big(S_2(\bar C,C)+\widetilde\tr \bar\eta^jC^j +\bar C^j\widetilde{\eta}^j \big)}\nonumber\\
&=&\exp\big(\sum\widetilde{\bar\eta}^{j_1}_{mn}P^{j_1 j_2}_{mn;kl}\widetilde{\eta}^{j_2}_{kl}\big)\label{functionalfreeghost}
\end{eqnarray}
where $S_2(\bar C,C)$ given by \eqref{quadrat-ghost} and $\eta$ and $\bar \eta$ with respective ghost number $+1$ and $-1$ denote respectively the sources for $\bar C$ and $C$.  The tilde over the  fields denotes as before the redefinition of the sources by a factor of $k(j+1)$. Correlation functions involving ghost fields are obtained as usual by the  action of functional derivatives $\frac{\delta}{\delta{\widetilde{\bar\eta}}_{nm}}$ and $\frac{\delta}{\delta\widetilde \eta_{nm}}$ with left and right action defined by the following generic relations
\begin{eqnarray}
\frac{\delta}{\delta\widetilde{\bar\eta}^j_{nm}}\exp(\sum_{j,m,n}(\widetilde{\bar \eta}^j_{nm}C^j_{mn}+\bar C^j_{mn}\widetilde{\eta}_{nm}))&=&C^j_{mn}\exp(\sum_{j,m,n}(\widetilde{\bar\eta}^j_{nm}C^j_{mn}+\bar C^j_{mn}\widetilde{\eta}_{nm}))\label{derfunctetabar} \\
\frac{\delta}{\delta\widetilde{\eta}^j_{nm}}\exp(\sum_{j,m,n}(\widetilde{\bar\eta}^j_{nm}C^j_{mn}+\bar C^j_{mn}\widetilde{\eta}_{nm}))&=&\exp(\sum_{j,m,n}(\widetilde{\bar\eta}^j_{nm}C^j_{mn}+\bar C^j_{mn}\widetilde{\eta}_{nm}))\bar C^j_{mn}\label{derfuncteta}.
\end{eqnarray}
Recall that $\frac{\delta}{\delta\widetilde{\bar\eta}^j_{nm}}$ and $\frac{\delta}{\delta\widetilde{\eta}^j_{nm}}$ inherit respective ghost numbers of $\widetilde{\bar\eta}^j_{nm}$ and $\widetilde{\eta}^j_{nm}$ so that they commute (resp. anticommute) with objects with zero (resp. $+1$) ghost number, modulo 2. In particular, these functional derivatives obeys a graded Leibnitz rule, namely
\begin{equation}
\frac{\delta}{\delta\widetilde{\eta}^j_{nm}}(ab)=\frac{\delta}{\delta\widetilde{\eta}^j_{nm}}(a)b+(-1)^{\vert a\vert}a\frac{\delta}{\delta\widetilde{\eta}^j_{nm}}(b)
\end{equation}
where $\vert a\vert$ is the ghost number of $a$ (and similar rule for $\frac{\delta}{\delta\widetilde{\bar\eta}^j_{nm}}$).

The generating functional of the connected Green functions $W(J,\eta,\bar\eta)$ is defined by
\begin{eqnarray}
Z(J,\eta,\bar\eta)&=&\exp(W(J,\eta,\bar\eta)
={\cal{N}}\exp(-S_{int}(\frac{\delta}{\delta \widetilde J},\frac{\delta}{\delta\widetilde{\bar\eta}},\frac{\delta}{\delta\widetilde{\eta}}))\nonumber\\
&\times&\exp\big(\frac{1}{8}\sum(\widetilde J_i)^{j_1}_{mn}P^{j_1 j_2}_{mn;kl} (\widetilde J_i)^{j_2}_{kl}\big)\exp\big(\sum\widetilde{\bar\eta}^{j_1}_{mn}P^{j_1 j_2}_{mn;kl}\widetilde{\eta}^{j_2}_{kl}\big)\label{functionalconnected}
\end{eqnarray}
where ${\cal{N}}$ is an unessential pre-factor and the interaction factor $S_{int}(\frac{\delta}{\delta \widetilde J},\frac{\delta}{\delta\widetilde{\bar\eta}},\frac{\delta}{\delta\widetilde{\eta}})$ can be read off from the gauge fixed action \eqref{gaugefixfinal}. 

It is convenient to focus on the effective action $\Gamma(A, \bar C, C)$ related to $W(J,\eta,\bar\eta)$ \eqref{functionalconnected} by the following Legendre transform 
\begin{equation}
\Gamma(A, \bar C, C)=\sum_{j,m,n} (A_i)^j_{mn}(\widetilde J_i)^j_{nm}+\widetilde{\bar\eta}^j_{nm}C^j_{mn}+\bar C^j_{mn}\widetilde{\eta}_{nm}-W(J,\eta,\bar\eta)\label{legendre1},
\end{equation}
\begin{equation}
(A_i)^j_{mn}=\frac{\delta W(J,\eta,\bar\eta)}{\delta(\widetilde J_i)^j_{nm}};\;\ \bar C^j_{mn}=\frac{\delta W(J,\eta,\bar\eta)}{\delta\widetilde\eta^j_{nm}};\; C^j_{mn}=\frac{\delta W(J,\eta,\bar\eta)}{\delta\widetilde{\bar\eta}^j_{nm}}\label{legendre2}.
\end{equation}
We are now in position to examine the fate of the one-point function for the gauge potential at the 1-loop order stemming from the 3-linear vertices. \par 

It is convenient to consider the perturbative expansion obtained from
\begin{equation}
W(J,\eta,\bar\eta)=W_0(J,\eta,\bar\eta)+\ln\big(1+e^{-W_0(J,\eta,\bar\eta)}[e^{-S_{int}(\frac{\delta}{\delta \widetilde J},\frac{\delta}{\delta\widetilde{\bar\eta}},\frac{\delta}{\delta\widetilde{\eta}})}-1]e^{W_0(J,\eta,\bar\eta)} \big)\label{formalW},
\end{equation}
\begin{equation}
W_0(J,\eta,\bar\eta):=\frac{1}{8}\sum(\widetilde J_i)^{j_1}_{mn}P^{j_1 j_2}_{mn;kl} (\widetilde J_i)^{j_2}_{kl}+\sum\widetilde{\bar\eta}^{j_1}_{mn}P^{j_1 j_2}_{mn;kl}\widetilde{\eta}^{j_2}_{kl}\label{formalW0},
\end{equation}
in view of \eqref{functionalconnected}, by further expanding the functional logarithm contribution in the RHS of \eqref{formalW} in which $e^{-S_{int}}$ is understood as a formal series in $S_{int}$ as usual{\footnote{We do not write explicitly the coupling constant. This latter can be easily restored}}. In the following, the expression for the terms contributing to our one-loop analysis reduces to
\begin{equation}
W(J,\eta,\bar\eta)=W_0(J,\eta,\bar\eta)-e^{-W_0(J,\eta,\bar\eta)}S_{int}(\frac{\delta}{\delta \widetilde J},\frac{\delta}{\delta\widetilde{\bar\eta}},\frac{\delta}{\delta\widetilde\eta})e^{W_0(J,\eta,\bar\eta)}+...\label{1loopgeneric}.
\end{equation}
This combined with \eqref{legendre1} yields the expression for the one-loop effective action by solving perturbatively \eqref{legendre2}; namely, one obtains
\begin{equation}
(A_i)^j_{mn}=\frac{1}{4}P_{nm;kl}(\widetilde J_i)_{kl}+...\ ;\ C^j_{mn}=P_{nm;kl}\widetilde\eta^j_{kl} +...\ ;\ \bar C^j_{mn}=\widetilde{\bar\eta}^j_{kl}P_{kl;nm} +...\label{legendreinter}
\end{equation}
where has been used \eqref{cons-indiceprim} (which implies $P_{mn;kl}=P_{kl;mn}$) to obtain the 1st relation and the dots  denote irrelevant higher order terms. Thus,
\begin{equation}
(\widetilde J_i)^j_{sr}=4\Delta_{rs;mn}(A_i)^j_{mn}+...\ ;\ \widetilde\eta^j_{sr}=\Delta_{rs;mn}C^j_{mn}+...\ ;\ \widetilde{\bar\eta}^j_{sr}=\bar C^j_{mn}\Delta_{mn;rs}+...\label{inverslegendre}
\end{equation}
\section{\texorpdfstring{Large $j$ limit of the tadpole}{Large j limit }}\label{appendix2}

To simplify the expressions, we set $\lambda=1$ through this appendix. By combining to \eqref{goldstoneoperator1}, the recursion formula (A.22) of ref \cite{vit-wal-12} relating the fuzzy spherical harmonics given by 
\begin{equation}
F^j(m)F^j(n)(Y^j_{lk})_{m+1,n+1}+F^j(-m)F^j(-n)(Y^j_{lk})_{m-1,n-1}=\Phi(j,l;mn)(Y^j_{lk})_{mn},
\end{equation}
where
\begin{equation}
\Phi(j,l;mn)=2j(j+1)-l(l+1)-2mn,\label{phi-interm}
\end{equation}
and further using
\begin{equation}
(Y^j_{lk})_{-m,n}=(-1)^{2j+l}(Y^j_{l,-k})_{m,-n},\ P_{-m-1,n-1;n,-m}=P_{m+1,-n+1;-n,m},
\end{equation}
together with \eqref{sum-boundary}, one obtains after some algebra
\begin{equation}
F^j(n)\Pi(j,n)+F^j(-n)\Pi(j,-n)=R(j,l;n),\forall j\in\frac{\mathbb{N}}{2},\ -j\le n\le j,\label{recu-pi}
\end{equation}
\begin{equation}
R(j;n)=2\sum_{m,l,k}\frac{(-1)^{2j}\Phi(j,l;mn)}{l(l+1)(2j+1)}(Y^j_{lk})^2_{mn}-2(F^{j2}(n)+F^{j2}(-n))\sum_{m=-j}^jP_{mn;nm}\label{RHS}.
\end{equation}
In the summation over $l$, it is of course understood that the IR singularities stemming from the propagator have balanced each other in the 2 terms involved in $\Pi(j,n)$. Using now orthogonality relations among Wigner $3j$ symbols in \eqref{RHS}, it can be realized that only two contributions in the first term of the RHS of \eqref{RHS} are non vanishing while the one depending on $mn$ (see \eqref{phi-interm}) vanishes thanks to the selection rules on the Wigner $3j$ symbols. The second term in \eqref{RHS} can be easily computed from \eqref{sum-boundary}. Finally, one obtains
\begin{equation}
R(j;n)=2(-1)^{2n}(\frac{2n^2}{2j+1}\sum_{l=1}^{2j}\frac{2l+1}{l(l+1)}-(2j+1)), \forall j\in\frac{\mathbb{N}}{2},\ -j\le n\le j\label{RHSfinal}.
\end{equation}
From \eqref{recu-pi} and \eqref{RHSfinal}, it follows that ($j\ne0$)
\begin{eqnarray}
\Pi(j,0)&=& -\frac{2j+1}{\sqrt{j(j+1)}}\label{pizero}\\
\Pi(j,-j)&=& \frac{(-1)^{2j}2}{{\sqrt{2j}}}(\frac{2j^2}{2j+1}\sum_{l=1}^{2j}\frac{2l+1}{l(l+1)}-(2j+1))\label{pij}.
\end{eqnarray}
Using the fact that $\lim_{j\to\infty}\sum_{l=1}^{2j}\frac{2l+1}{l(l+1)}=\int_1^\infty dx\frac{2x+1}{x(x+1)}$, one deduces from \eqref{pizero} and \eqref{pij} that
\begin{equation}
\lim_{j\to\infty}\Pi(j,0)=-2,\ \lim_{j\to\infty}\vert \Pi(j,-j)\vert=+\infty\label{limit-pi}
\end{equation}
where in particular one obtains $\Pi(j,-j)\sim{\sqrt{j}}\log j$ at large $j$.

\end{document}